\documentclass[letter,longauth]{aa}
\usepackage{graphicx}
\usepackage{txfonts}
\usepackage[colorlinks=true,citecolor=blue]{hyperref}
\usepackage{multirow}
\begin{document}

   \title{TOI-1055\,b: Neptunian planet characterised with HARPS, TESS, and CHEOPS\thanks{
   This article uses data from CHEOPS program CH\_PR100024.}\fnmsep
   \thanks{Based on observations made with ESO-3.6 m telescope at the La Silla Observatory under programme IDs 106.21TJ.001, 188.C-0265, and 0100.D-044.}}

   \author{
A. Bonfanti\inst{1} $^{\href{https://orcid.org/0000-0002-1916-5935}{\includegraphics[scale=0.5]{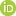}}}$\and
D. Gandolfi\inst{2} $^{\href{https://orcid.org/0000-0001-8627-9628}{\includegraphics[scale=0.5]{figures/orcid.jpg}}}$\and
J. A. Egger\inst{3} $^{\href{https://orcid.org/0000-0003-1628-4231}{\includegraphics[scale=0.5]{figures/orcid.jpg}}}$\and
L. Fossati\inst{1} $^{\href{https://orcid.org/0000-0003-4426-9530}{\includegraphics[scale=0.5]{figures/orcid.jpg}}}$\and
J. Cabrera\inst{4}\and
A. Krenn\inst{1} $^{\href{https://orcid.org/0000-0003-3615-4725}{\includegraphics[scale=0.5]{figures/orcid.jpg}}}$\and
Y. Alibert\inst{3} $^{\href{https://orcid.org/0000-0002-4644-8818}{\includegraphics[scale=0.5]{figures/orcid.jpg}}}$\and
W. Benz\inst{3,5} $^{\href{https://orcid.org/0000-0001-7896-6479}{\includegraphics[scale=0.5]{figures/orcid.jpg}}}$\and
N. Billot\inst{6} $^{\href{https://orcid.org/0000-0003-3429-3836}{\includegraphics[scale=0.5]{figures/orcid.jpg}}}$\and
H.-G. Florén\inst{7}\and
M. Lendl\inst{6} $^{\href{https://orcid.org/0000-0001-9699-1459}{\includegraphics[scale=0.5]{figures/orcid.jpg}}}$\and
V. Adibekyan\inst{8} $^{\href{https://orcid.org/0000-0002-0601-6199}{\includegraphics[scale=0.5]{figures/orcid.jpg}}}$\and
S. Salmon\inst{6} $^{\href{https://orcid.org/0000-0002-1714-3513}{\includegraphics[scale=0.5]{figures/orcid.jpg}}}$\and
N. C. Santos\inst{8,9} $^{\href{https://orcid.org/0000-0003-4422-2919}{\includegraphics[scale=0.5]{figures/orcid.jpg}}}$\and
S. G. Sousa\inst{8} $^{\href{https://orcid.org/0000-0001-9047-2965}{\includegraphics[scale=0.5]{figures/orcid.jpg}}}$\and
T. G. Wilson\inst{10} $^{\href{https://orcid.org/0000-0001-8749-1962}{\includegraphics[scale=0.5]{figures/orcid.jpg}}}$\and
O. Barragán\inst{11}\and
A. Collier Cameron\inst{10} $^{\href{https://orcid.org/0000-0002-8863-7828}{\includegraphics[scale=0.5]{figures/orcid.jpg}}}$\and
L. Delrez\inst{12,13} $^{\href{https://orcid.org/0000-0001-6108-4808}{\includegraphics[scale=0.5]{figures/orcid.jpg}}}$\and
M. Esposito\inst{14}\and
E. Goffo\inst{2,14}\and
H. Osborne\inst{15}\and
H. P. Osborn\inst{5,16} $^{\href{https://orcid.org/0000-0002-4047-4724}{\includegraphics[scale=0.5]{figures/orcid.jpg}}}$\and
L. M. Serrano\inst{2} $^{\href{https://orcid.org/0000-0001-9211-3691}{\includegraphics[scale=0.5]{figures/orcid.jpg}}}$\and
V. Van Eylen\inst{15}\and
J. Alarcon\inst{17}\and
R. Alonso\inst{18,19} $^{\href{https://orcid.org/0000-0001-8462-8126}{\includegraphics[scale=0.5]{figures/orcid.jpg}}}$\and
G. Anglada\inst{20,21} $^{\href{https://orcid.org/0000-0002-3645-5977}{\includegraphics[scale=0.5]{figures/orcid.jpg}}}$\and
T. Bárczy\inst{22} $^{\href{https://orcid.org/0000-0002-7822-4413}{\includegraphics[scale=0.5]{figures/orcid.jpg}}}$\and
D. Barrado Navascues\inst{23} $^{\href{https://orcid.org/0000-0002-5971-9242}{\includegraphics[scale=0.5]{figures/orcid.jpg}}}$\and
S. C. C. Barros\inst{8,9} $^{\href{https://orcid.org/0000-0003-2434-3625}{\includegraphics[scale=0.5]{figures/orcid.jpg}}}$\and
W. Baumjohann\inst{1} $^{\href{https://orcid.org/0000-0001-6271-0110}{\includegraphics[scale=0.5]{figures/orcid.jpg}}}$\and
M. Beck\inst{6} $^{\href{https://orcid.org/0000-0003-3926-0275}{\includegraphics[scale=0.5]{figures/orcid.jpg}}}$\and
T. Beck\inst{3}\and
M. Bedell\inst{24} $^{\href{https://orcid.org/0000-0001-9907-7742}{\includegraphics[scale=0.5]{figures/orcid.jpg}}}$\and
X. Bonfils\inst{25} $^{\href{https://orcid.org/0000-0001-9003-8894}{\includegraphics[scale=0.5]{figures/orcid.jpg}}}$\and
L. Borsato\inst{26} $^{\href{https://orcid.org/0000-0003-0066-9268}{\includegraphics[scale=0.5]{figures/orcid.jpg}}}$\and
A. Brandeker\inst{7} $^{\href{https://orcid.org/0000-0002-7201-7536}{\includegraphics[scale=0.5]{figures/orcid.jpg}}}$\and
C. Broeg\inst{3,27} $^{\href{https://orcid.org/0000-0001-5132-2614}{\includegraphics[scale=0.5]{figures/orcid.jpg}}}$\and
S. Charnoz\inst{28} $^{\href{https://orcid.org/0000-0002-7442-491X}{\includegraphics[scale=0.5]{figures/orcid.jpg}}}$\and
C. Corral Van Damme\inst{29}\and
Sz. Csizmadia\inst{4} $^{\href{https://orcid.org/0000-0001-6803-9698}{\includegraphics[scale=0.5]{figures/orcid.jpg}}}$\and
P. E. Cubillos\inst{30,1}\and
M. B. Davies\inst{31} $^{\href{https://orcid.org/0000-0001-6080-1190}{\includegraphics[scale=0.5]{figures/orcid.jpg}}}$\and
M. Deleuil\inst{32} $^{\href{https://orcid.org/0000-0001-6036-0225}{\includegraphics[scale=0.5]{figures/orcid.jpg}}}$\and
O. D. S. Demangeon\inst{8,9} $^{\href{https://orcid.org/0000-0001-7918-0355}{\includegraphics[scale=0.5]{figures/orcid.jpg}}}$\and
B.-O. Demory\inst{5,3} $^{\href{https://orcid.org/0000-0002-9355-5165}{\includegraphics[scale=0.5]{figures/orcid.jpg}}}$\and
D. Ehrenreich\inst{6,33} $^{\href{https://orcid.org/0000-0001-9704-5405}{\includegraphics[scale=0.5]{figures/orcid.jpg}}}$\and
A. Erikson\inst{4}\and
A. Fortier\inst{3,5} $^{\href{https://orcid.org/0000-0001-8450-3374}{\includegraphics[scale=0.5]{figures/orcid.jpg}}}$\and
M. Fridlund\inst{34,35} $^{\href{https://orcid.org/0000-0002-0855-8426}{\includegraphics[scale=0.5]{figures/orcid.jpg}}}$\and
M. Gillon\inst{12} $^{\href{https://orcid.org/0000-0003-1462-7739}{\includegraphics[scale=0.5]{figures/orcid.jpg}}}$\and
M. Güdel\inst{36}\and
S. Hoyer\inst{32} $^{\href{https://orcid.org/0000-0003-3477-2466}{\includegraphics[scale=0.5]{figures/orcid.jpg}}}$\and
K. G. Isaak\inst{37} $^{\href{https://orcid.org/0000-0001-8585-1717}{\includegraphics[scale=0.5]{figures/orcid.jpg}}}$\and
F. Kerschbaum\inst{38}\and
L. L. Kiss\inst{39,40}\and
J. Laskar\inst{41} $^{\href{https://orcid.org/0000-0003-2634-789X}{\includegraphics[scale=0.5]{figures/orcid.jpg}}}$\and
A. Lecavelier des Etangs\inst{42} $^{\href{https://orcid.org/0000-0002-5637-5253}{\includegraphics[scale=0.5]{figures/orcid.jpg}}}$\and
D. Lorenzo-Oliveira\inst{43,44} $^{\href{https://orcid.org/0000-0002-1387-2954}{\includegraphics[scale=0.5]{figures/orcid.jpg}}}$\and
C. Lovis\inst{6} $^{\href{https://orcid.org/0000-0001-7120-5837}{\includegraphics[scale=0.5]{figures/orcid.jpg}}}$\and
D. Magrin\inst{26} $^{\href{https://orcid.org/0000-0003-0312-313X}{\includegraphics[scale=0.5]{figures/orcid.jpg}}}$\and
L. Marafatto\inst{26} $^{\href{https://orcid.org/0000-0002-8822-6834}{\includegraphics[scale=0.5]{figures/orcid.jpg}}}$\and
P. F. L. Maxted\inst{45} $^{\href{https://orcid.org/0000-0003-3794-1317}{\includegraphics[scale=0.5]{figures/orcid.jpg}}}$\and
J. Mel\'endez\inst{44}\and
C. Mordasini\inst{46,5}\and
V. Nascimbeni\inst{26} $^{\href{https://orcid.org/0000-0001-9770-1214}{\includegraphics[scale=0.5]{figures/orcid.jpg}}}$\and
G. Olofsson\inst{7} $^{\href{https://orcid.org/0000-0003-3747-7120}{\includegraphics[scale=0.5]{figures/orcid.jpg}}}$\and
R. Ottensamer\inst{38}\and
I. Pagano\inst{47} $^{\href{https://orcid.org/0000-0001-9573-4928}{\includegraphics[scale=0.5]{figures/orcid.jpg}}}$\and
E. Pallé\inst{18} $^{\href{https://orcid.org/0000-0003-0987-1593}{\includegraphics[scale=0.5]{figures/orcid.jpg}}}$\and
G. Peter\inst{48} $^{\href{https://orcid.org/0000-0001-6101-2513}{\includegraphics[scale=0.5]{figures/orcid.jpg}}}$\and
D. Piazza\inst{3}\and
G. Piotto\inst{26,49} $^{\href{https://orcid.org/0000-0002-9937-6387}{\includegraphics[scale=0.5]{figures/orcid.jpg}}}$\and
D. Pollacco\inst{50}\and
D. Queloz\inst{51,52} $^{\href{https://orcid.org/0000-0002-3012-0316}{\includegraphics[scale=0.5]{figures/orcid.jpg}}}$\and
R. Ragazzoni\inst{26,49} $^{\href{https://orcid.org/0000-0002-7697-5555}{\includegraphics[scale=0.5]{figures/orcid.jpg}}}$\and
N. Rando\inst{29}\and
H. Rauer\inst{4,53,54} $^{\href{https://orcid.org/0000-0002-6510-1828}{\includegraphics[scale=0.5]{figures/orcid.jpg}}}$\and
I. Ribas\inst{20,21} $^{\href{https://orcid.org/0000-0002-6689-0312}{\includegraphics[scale=0.5]{figures/orcid.jpg}}}$\and
G. Scandariato\inst{47} $^{\href{https://orcid.org/0000-0003-2029-0626}{\includegraphics[scale=0.5]{figures/orcid.jpg}}}$\and
D. Ségransan\inst{6} $^{\href{https://orcid.org/0000-0003-2355-8034}{\includegraphics[scale=0.5]{figures/orcid.jpg}}}$\and
A. E. Simon\inst{3} $^{\href{https://orcid.org/0000-0001-9773-2600}{\includegraphics[scale=0.5]{figures/orcid.jpg}}}$\and
A. M. S. Smith\inst{4} $^{\href{https://orcid.org/0000-0002-2386-4341}{\includegraphics[scale=0.5]{figures/orcid.jpg}}}$\and
M. Steller\inst{1} $^{\href{https://orcid.org/0000-0003-2459-6155}{\includegraphics[scale=0.5]{figures/orcid.jpg}}}$\and
Gy. M. Szabó\inst{55,56}\and
N. Thomas\inst{3}\and
S. Udry\inst{6} $^{\href{https://orcid.org/0000-0001-7576-6236}{\includegraphics[scale=0.5]{figures/orcid.jpg}}}$\and
B. Ulmer\inst{48}\and
V. Van Grootel\inst{13} $^{\href{https://orcid.org/0000-0003-2144-4316}{\includegraphics[scale=0.5]{figures/orcid.jpg}}}$\and
J. Venturini\inst{6}\and
N. A. Walton\inst{57} $^{\href{https://orcid.org/0000-0003-3983-8778}{\includegraphics[scale=0.5]{figures/orcid.jpg}}}$
          }

   \institute{\label{inst:1} Space Research Institute, Austrian Academy of Sciences, Schmiedlstrasse 6, A-8042 Graz, Austria \\
              \email{andrea.bonfanti@oeaw.ac.at}
         \and
\label{inst:2} Dipartimento di Fisica, Universita degli Studi di Torino, via Pietro Giuria 1, I-10125, Torino, Italy \and
\label{inst:3} Physikalisches Institut, University of Bern, Sidlerstrasse 5, 3012 Bern, Switzerland \and
\label{inst:4} Institute of Planetary Research, German Aerospace Center (DLR), Rutherfordstrasse 2, 12489 Berlin, Germany \and
\label{inst:5} Center for Space and Habitability, University of Bern, Gesellschaftsstrasse 6, 3012 Bern, Switzerland \and
\label{inst:6} Observatoire Astronomique de l'Université de Genève, Chemin Pegasi 51, CH-1290 Versoix, Switzerland \and
\label{inst:7} Department of Astronomy, Stockholm University, AlbaNova University Center, 10691 Stockholm, Sweden \and
\label{inst:8} Instituto de Astrofisica e Ciencias do Espaco, Universidade do Porto, CAUP, Rua das Estrelas, 4150-762 Porto, Portugal \and
\label{inst:9} Departamento de Fisica e Astronomia, Faculdade de Ciencias, Universidade do Porto, Rua do Campo Alegre, 4169-007 Porto, Portugal \and
\label{inst:10} Centre for Exoplanet Science, SUPA School of Physics and Astronomy, University of St Andrews, North Haugh, St Andrews KY16 9SS, UK \and
\label{inst:11} Sub-department of Astrophysics, Department of Physics, University of Oxford, Oxford, OX1 3RH, UK \and
\label{inst:12} Astrobiology Research Unit, Université de Liège, Allée du 6 Août 19C, B-4000 Liège, Belgium \and
\label{inst:13} Space sciences, Technologies and Astrophysics Research (STAR) Institute, Université de Liège, Allée du 6 Août 19C, 4000 Liège, Belgium \and
\label{inst:14} Thüringer Landessternwarte Tautenburg, Sternwarte 5, D-07778 Tautenburg, Germany \and
\label{inst:15} Mullard Space Science Laboratory, University College London,Holmbury St. Mary, Dorking, Surrey, RH5 6NT, UK \and
\label{inst:16} Department of Physics and Kavli Institute for Astrophysics and Space Research, Massachusetts Institute of Technology, Cambridge, MA 02139, USA \and
\label{inst:17} European Southern Observatory, Alonso de Cordova 3107, Vitacura, Santiago de Chile, Chile \and
\label{inst:18} Instituto de Astrofisica de Canarias, 38200 La Laguna, Tenerife, Spain \and
\label{inst:19} Departamento de Astrofisica, Universidad de La Laguna, 38206 La Laguna, Tenerife, Spain \and
\label{inst:20} Institut de Ciencies de l'Espai (ICE, CSIC), Campus UAB, Can Magrans s/n, 08193 Bellaterra, Spain \and
\label{inst:21} Institut d'Estudis Espacials de Catalunya (IEEC), 08034 Barcelona, Spain \and
\label{inst:22} Admatis, 5. Kandó Kálmán Street, 3534 Miskolc, Hungary \and
\label{inst:23} Depto. de Astrofisica, Centro de Astrobiologia (CSIC-INTA), ESAC campus, 28692 Villanueva de la Cañada (Madrid), Spain \and
\label{inst:24} Center for Computational Astrophysics, Flatiron Institute, New York, NY \and
\label{inst:25} Université Grenoble Alpes, CNRS, IPAG, 38000 Grenoble, France \and
\label{inst:26} INAF, Osservatorio Astronomico di Padova, Vicolo dell'Osservatorio 5, 35122 Padova, Italy \and
\label{inst:27} Center for Space and Habitability, Gesellsschaftstrasse 6, 3012 Bern, Switzerland \and
\label{inst:28} Université de Paris, Institut de physique du globe de Paris, CNRS, F-75005 Paris, France \and
\label{inst:29} ESTEC, European Space Agency, 2201AZ, Noordwijk, NL \and
\label{inst:30} INAF, Osservatorio Astrofisico di Torino, Via Osservatorio, 20, I-10025 Pino Torinese To, Italy \and
\label{inst:31} Centre for Mathematical Sciences, Lund University, Box 118, 221 00 Lund, Sweden \and
\label{inst:32} Aix Marseille Univ, CNRS, CNES, LAM, 38 rue Frédéric Joliot-Curie, 13388 Marseille, France \and
\label{inst:33} Centre Vie dans l’Univers, Faculté des sciences, Université de Genève, Quai Ernest-Ansermet 30, CH-1211 Gen\`eve 4, Switzerland \and
\label{inst:34} Leiden Observatory, University of Leiden, PO Box 9513, 2300 RA Leiden, The Netherlands \and
\label{inst:35} Department of Space, Earth and Environment, Chalmers University of Technology, Onsala Space Observatory, 439 92 Onsala, Sweden \and
\label{inst:36} University of Vienna, Department of Astrophysics, Türkenschanzstrasse 17, 1180 Vienna, Austria \and
\label{inst:37} Science and Operations Department - Science Division (SCI-SC), Directorate of Science, European Space Agency (ESA), European Space Research and Technology Centre (ESTEC),
Keplerlaan 1, 2201-AZ Noordwijk, The Netherlands \and
\label{inst:38} Department of Astrophysics, University of Vienna, Tuerkenschanzstrasse 17, 1180 Vienna, Austria \and
\label{inst:39} Konkoly Observatory, Research Centre for Astronomy and Earth Sciences, 1121 Budapest, Konkoly Thege Miklós út 15-17, Hungary \and
\label{inst:40} ELTE E\"otv\"os Lor\'and University, Institute of Physics, P\'azm\'any P\'eter s\'et\'any 1/A, 1117 Budapest, Hungary \and
\label{inst:41} IMCCE, UMR8028 CNRS, Observatoire de Paris, PSL Univ., Sorbonne Univ., 77 av. Denfert-Rochereau, 75014 Paris, France \and
\label{inst:42} Institut d'astrophysique de Paris, UMR7095 CNRS, Université Pierre \& Marie Curie, 98bis blvd. Arago, 75014 Paris, France \and
\label{inst:43} Laborat\'{o}rio Nacional de Astrof\'{i}sica, Rua Estados Unidos 154, 37504-364, Itajub\'{a} - MG, Brazil \and
\label{inst:44} Universidade de S\~{a}o Paulo, IAG, Departamento de Astronomia, Rua do Mat\~{a}o 1226, 05509-900, S\~{a}o Paulo, SP, Brazil \and
\label{inst:45} Astrophysics Group, Keele University, Staffordshire, ST5 5BG, United Kingdom \and
\label{inst:46} Physikalisches Institut, University of Bern, Gesellschaftsstrasse 6, 3012 Bern, Switzerland \and
\label{inst:47} INAF, Osservatorio Astrofisico di Catania, Via S. Sofia 78, 95123 Catania, Italy \and
\label{inst:48} Institute of Optical Sensor Systems, German Aerospace Center (DLR), Rutherfordstrasse 2, 12489 Berlin, Germany \and
\label{inst:49} Dipartimento di Fisica e Astronomia "Galileo Galilei", Universita degli Studi di Padova, Vicolo dell'Osservatorio 3, 35122 Padova, Italy \and
\label{inst:50} Department of Physics, University of Warwick, Gibbet Hill Road, Coventry CV4 7AL, United Kingdom \and
\label{inst:51} ETH Zurich, Department of Physics, Wolfgang-Pauli-Strasse 2, CH-8093 Zurich, Switzerland \and
\label{inst:52} Cavendish Laboratory, JJ Thomson Avenue, Cambridge CB3 0HE, UK \and
\label{inst:53} Zentrum für Astronomie und Astrophysik, Technische Universität Berlin, Hardenbergstr. 36, D-10623 Berlin, Germany \and
\label{inst:54} Institut für Geologische Wissenschaften, Freie Universität Berlin, 12249 Berlin, Germany \and
\label{inst:55} ELTE E\"otv\"os Lor\'and University, Gothard Astrophysical Observatory, 9700 Szombathely, Szent Imre h. u. 112, Hungary \and
\label{inst:56} MTA-ELTE Exoplanet Research Group, 9700 Szombathely, Szent Imre h. u. 112, Hungary \and
\label{inst:57} Institute of Astronomy, University of Cambridge, Madingley Road, Cambridge, CB3 0HA, United Kingdom
             }

   \date{}

 
  \abstract
   {TOI-1055 is a Sun-like star known to host a transiting Neptune-sized planet on a 17.5-day orbit (TOI-1055\,b). Radial velocity (RV) analyses carried out by two independent groups using nearly the same set of HARPS spectra have provided measurements of planetary masses that differ by $\sim$\,2$\sigma$.}
   {Our aim in this work is to solve the inconsistency in the published planetary masses by significantly extending the set of HARPS RV measurements and employing a new analysis tool that is able to account and correct for stellar activity. Our further aim was to improve the precision on measurements of the planetary radius by observing two transits of the planet with the CHEOPS space telescope.}
   {We fit a skew normal (SN) function to each cross correlation function extracted from the HARPS spectra to obtain RV measurements and hyperparameters to be used for the detrending. We evaluated the correlation changes of the hyperparameters along the RV time series using the breakpoint technique. We performed a joint photometric and RV analysis using a Markov chain Monte Carlo (MCMC) scheme to simultaneously detrend the light curves and the RV time series.}
   {We firmly detected the Keplerian signal of TOI-1055\,b, deriving a planetary mass of $M_b=20.4_{-2.5}^{+2.6}\,M_{\oplus}$ ($\sim$12\%). This value is in agreement with one of the two estimates in the literature, but it is significantly more precise. Thanks to the TESS transit light curves combined with exquisite CHEOPS photometry, we also derived a planetary radius of $R_b=3.490_{-0.064}^{+0.070}\,R_{\oplus}$ ($\sim$1.9\%). Our mass and radius measurements imply a mean density of $\rho_b=2.65_{-0.35}^{+0.37}$ g\,cm$^{-3}$ ($\sim$14\%). We further inferred the planetary structure and found that TOI-1055\,b is very likely to host a substantial gas envelope with a mass of $0.41^{+0.34}_{-0.20}$ M$_\oplus$ and a thickness of $1.05^{+0.30}_{-0.29}$ R$_\oplus$.}
   {Our RV extraction combined with the breakpoint technique has played a key role in the optimal removal of stellar activity from the HARPS time series, enabling us to solve the tension in the planetary mass values published so far for TOI-1055\,b.}


   \maketitle
%
%
\section{Introduction}
TOI-1055 (also known as HD\,183579) is a bright (V=8.68) Sun-like star hosting a transiting Neptune-sized planet on a $\sim$17.5-day orbit that was discovered by the Transiting Exoplanet Survey Satellite \citep[TESS;][]{ricker2015}. Recently, \citet[][hereafter P21]{palatnick2021} and \citet[][hereafter G21]{gan2021} carried out two independent radial velocity (RV) analyses of the TOI-1055 system, which led to a tension at about the 2$\sigma$ level with regard to the derived semi-amplitude $K_\mathrm{b}$ of the Doppler reflex motion induced by the transiting planet.

In more detail, while searching for archival RV data to possibly confirm exoplanet candidates among the TESS Objects of Interest \citep[TOIs;][]{Guerrero2021}, P21 found 53 archival measurements of TOI-1055 acquired between 13 October 2011 and 21 October 2017 (UT) with the High Accuracy Radial velocity Planet Searcher \citep[HARPS;][]{Mayor2003} spectrograph mounted at the European Southern Observatory (ESO) 3.6 m telescope. They found that the 53 HARPS RVs are in phase with the TESS transit ephemeris and consistent with a semi-amplitude variation of $K_\mathrm{b,P21}=4.9_{-1.0}^{+0.9}$ m\,s$^{-1}$. After determining the isochronal stellar parameters and performing a photometric analysis of the three available TESS transits, P21 derived a planetary mass of $M_{\rm b,P21}=19.7_{-3.9}^{+4.0}\,M_{\oplus}$, a radius of $R_{\rm b,P21}=3.55_{-0.12}^{+0.15}\,R_{\oplus}$, and a mean density of $\rho_{\rm b,P21}=2.39_{-0.54}^{+0.57}$ g\,cm$^{-3}$.

G21 carried out an independent analysis of the TOI-1055 system, modelling the same TESS data used by P21, but further including two partial ground-based transit light curves acquired with the Las Cumbres Observatory Global Telescope \citep[LCOGT;][]{brown2013}.
The results obtained by G21 from the transit modelling and stellar characterisation are consistent with those given by P21. In addition to the 53 HARPS RVs used by P21, G21 benefitted from three further HARPS spectra taken between 16 and 20 August 2019. When analysing the subset of 53 RV data only, G21 obtained results consistent with P21. However, the analysis performed on the complete set of 56 HARPS RVs provided a $K_{\mathrm b}$ value lower by a factor of 2, which differs by $\sim$1.7$\sigma$ from the P21 estimate. This implies an equal decrease in the planetary mass, $M_{\rm b}$, and density, $\rho_{\rm b}$. Taking into account these results, G21 finally adopted a mass of $M_{\rm b,G21}=11.2\pm5.4\,M_{\oplus}$ and a mean density of $\rho_{\rm b,G21}=1.4^{+0.9}_{-0.8}$ g\,cm$^{-3}$, remarking that their planetary mass measurement is not statistically significant being at the 2.1$\sigma$ level.

To ease the tension in the measured planetary mass, we refined the stellar parameters and performed a transit and RV joint analysis, employing new transit light curves (LCs) obtained with the Characterising Exoplanet Satellite \citep[CHEOPS;][]{benz2020}, along with additional HARPS RVs. The Doppler measurements were extracted and detrended applying a Skew Normal (SN) fit \citep{simola2019} onto the cross-correlation functions (CCFs) combined with the breakpoint (\emph{bp}) algorithm \citep{simola2022}, a new technique specifically developed to disentangle stellar activity from the Doppler reflex motion induced by orbiting planets. This allows us to remove the RV contribution arising from stellar variability and derive more precise planetary parameters, enabling planetary interior modelling.

This paper is organised as follows. In Sect.~\ref{sec:data}, we present the photometric and spectroscopic data. In Sect.~\ref{sec:star}, we give the updated stellar properties. In Sect.~\ref{sec:method}, we describe the algorithms and methods we used to analyse the data and perform the LC and RV joint analysis. Our results are presented in Sect.~\ref{sec:results}. Finally, in Sect.~\ref{sec:conclusions}, we summarise our work, discuss the results, and present our conclusions. 

\section{Observational data}\label{sec:data}
For the LC analysis, we used the three available TESS transits (also analysed by P21 and G21) complemented with two additional transit LCs acquired with the CHEOPS space telescope. We further expanded the available HARPS data-set comprising 56 spectra (also used by P21 and G21) with 16 additional HARPS spectra.
\subsection{TESS photometry}\label{sec:TESS}
TESS observed two transits of TOI-1055\,b in Sector 13 (19 June\,--\,17 July 2019; 20,479 data points) and a third one in Sector 27 (5\,--\,30 July 2020; 17,546 data points). We downloaded the data from the Mikulski Archive for Space Telescopes (MAST)\footnote{\url{https://mast.stsci.edu/portal/Mashup/Clients/Mast/Portal.html}} portal. For our analysis we used the Presearch Data Conditioning Simple Aperture Photometry \citep[PDCSAP;][]{Smith2012,Stumpe2012}, which is corrected for instrumental systematics, as processed by the TESS Science Processing Operation Center \citep[SPOC;][]{jenkins2016}. We discarded all data points marked with bad quality flags and applied a five-median absolute deviation (MAD) clipping algorithm to remove additional outliers. To save computational time while properly detrending the LCs, we extracted $\sim$12 hours of TESS data points centred around each transit.

\subsection{CHEOPS photometry}
The CHEOPS space telescope carried out two transit observations of TOI-1055\,b, the first on 25 May 2021 (file key: \texttt{CH\_PR100024\_TG012301\_V0200}) and the second one on 11 June 2021 (file key: \texttt{CH\_PR100024\_TG012302\_V0200}). CHEOPS photometry was processed using the dedicated data reduction pipeline \citep[version 13.1;][]{hoyer2020}, and is available via the Data \& Analysis Center for Exoplanets (DACE) database\footnote{\url{https://dace.unige.ch/cheopsDatabase/?}}. We discarded data points with bad quality flags and applied a 5-MAD-clipping to reject outliers; the raw CHEOPS LCs are shown in Fig.~\ref{fig:rawCheopsLC}. To identify and correct for systematics affecting CHEOPS LCs, the data also contain vectors reporting the temporal evolution of the telescope roll angle \citep[\texttt{roll}, for a complete discussion see][]{bonfanti2021}, flux of background stars inside the photometric aperture (\texttt{contamination}), smearing effect (\texttt{smear}, seen as trails on the CCD), background light (\texttt{bg}, e.g. due to zodiacal light), and the $x$- and $y$-coordinates of the point spread function centroid (\texttt{x}, \texttt{y}).

\subsection{HARPS high-resolution spectroscopy}\label{sec:HARPS}
We retrieved the 56 publicly available high-resolution (R\,$\approx$\,115,000) HARPS spectra of TOI-1055 from the ESO archive (prog. ID 188.C-0265, PI: J. Meléndez; prog. ID 0100.D-0444, PI: D. Lorenzo-Oliveira) and acquired 16 additional HARPS spectra between 3 October 2021 and 17 March 2022, as part of the observing programme 106.21TJ.001 (PI: D.\,Gandolfi). We reduced the 72 HARPS spectra using the dedicated HARPS data reduction software \citep[DRS][]{Lovis2007} and computed the cross-correlation function (CCF) using a G2 numerical mask \citep{Baranne1996,Pepe2002}. We also extracted the Ca\,{\sc ii} H\,\&\,K lines activity index log\,$R^\prime_\mathrm{HK}$ using the DRS, along with the star's color index B$-$V\,=\,0.653. In June 2015, the HARPS circular fibres were replaced with octagonal ones \citep{LoCurto2015}. We treated the HARPS RV measurements taken before (HARPS-pre, 36 measurements) and after (HARPS-post, 36 measurements) June 2015 as two independent data sets to account for the RV offset caused by the instrument upgrade.

\section{Host star characterisation}\label{sec:star}
\subsection{Spectroscopic and isochronal parameters}
TOI-1055 is a G2\,V star (P21) with a Tycho V-band magnitude of $V=8.68$ \citep{hog2000}, lying at a distance of $\sim$57\,pc from the Sun \citep{Gaia2021}. To derive the stellar spectroscopic parameters (i.e. effective temperature, $T_{\mathrm{eff}}$, metallicity, [Fe/H], and surface gravity, $\log{g}$), we co-added the HARPS spectra (see Sect.~\ref{sec:HARPS}). To this end, we applied the two-steps methodology described in \citet{santos2013} and \citet{sousa2021}. In brief, we first extracted the equivalent widths of the Fe lines listed in \citet{sousa2008}, employing the ARES code\footnote{The latest version of ARES code (v2) is available at \url{https://github.com/sousasag/ARES}.} \citep{sousa2007,sousa2015}. We then applied the MOOG code\footnote{\url{https://www.as.utexas.edu/~chris/moog.html}} \citep{sneden1973} to determine the atmospheric parameters using a grid of stellar atmospheric models from \citet{kurucz1993}. We thus obtained $T_{\mathrm{eff}}=5784\pm61$ K, $\mathrm{[Fe/H]}=-0.038\pm0.040$, and $\log{g}=4.45\pm0.10$ (cgs).

To derive the stellar radius, $R_{\star}$, we used the spectroscopic stellar parameters and the Gaia EDR3 parallax $\pi$ \citep{Gaia2021}, and we applied the infrared flux method \citep[IRFM][]{blackwell1977,schanche2020} via a Markov chain Monte Carlo (MCMC) algorithm. Within this framework, we compared the synthetic spectral energy distribution (SED) built from the catalogue of \textsc{ATLAS} models by \citet{castelli2003} with multi-photometry, as detailed in \citet{wilson2022}, deriving a stellar angular diameter $\theta=0.1576\pm0.0013$\,mas and obtaining $R_{\star}=0.960\pm0.006\,R_{\odot}$.

Finally, we determined the stellar mass, $M_\star$, and age, $t_\star$, using the derived $T_{\mathrm{eff}}$, [Fe/H], and $R_{\star}$ values, with their uncertainties, and two different sets of stellar evolutionary models, to improve the robustness of the results. In particular, we used the isochrone placement algorithm developed by \citet{bonfanti2015,bonfanti2016} to interpolate within pre-computed grids of PARSEC\footnote{\textsl{PA}dova and T\textsl{R}ieste \textsl{S}tellar \textsl{E}volutionary \textsl{C}ode: \url{http://stev.oapd.inaf.it/cgi-bin/cmd}.} v1.2S \citep{marigo2017} isochrones and tracks, and the Code Liègeois d'Évolution Stellaire \citep[CLES,][]{scuflaire2008} to build the best fitting evolutionary track according to the Levenberg-Marquadt minimisation scheme outlined in \citet{salmon2021}. We obtained $M_{\star}=0.996\pm0.049\,M_{\odot}$ and $t_{\star}=3.1\pm1.9$\,Gyr.
We note that the mean stellar density inferred from the isochronal mass and the IRFM radius, that is $\rho_{\star}=1.126\pm0.059\,\rho_{\odot}$, agrees well within 1$\sigma$ with its counterpart computed from the spectroscopic $\log{g}$ and the IRFM radius, that is $\rho_{\star,\mathrm{spec}}=1.09\pm0.25\,\rho_{\odot}$.

We refer the reader to \citet{bonfanti2021} for a detailed discussion of the theoretical models and statistical methodologies. The stellar parameters are listed in Table~\ref{tab:star} and are consistent within 1$\sigma$ with those found by P21 and G21.

\subsection{Rotation period}

Based on the stellar radius, $R_\star$, and projected rotational velocity, $v_\mathrm{rot}$\,sin\,$i_\star$, and assuming the star is seen equator-on ($i_\star$\,=\,90\,deg), G21 estimated the rotation period of TOI-1055 to be $P_\mathrm{rot,G21}$\,=\,24.6\,$\pm$\,2.6\,d. We independently inferred the rotation period from the relationship between log\,$R^\prime_\mathrm{HK}$ and $P_\mathrm{rot}$ empirically found by \citet{SuarezMascareno2015}. Using the mean chromospheric activity index log\,$R^\prime_\mathrm{HK}$\,=\, $-4.84\,\pm\,0.03$, as extracted from the HARPS spectra, we estimated a rotation period of $P_\mathrm{rot}$\,=\,$24.8_{-4.4}^{+5.4}$\,d in very good agreement with the value reported by G21. The empirical relation between stellar jitter and log\,$R^\prime_\mathrm{HK}$ from \citet{Hojjatpanah2020} predicts an activity-induced RV jitter of $\sim$\,4\,m\,s$^{-1}$.

\begin{table}
\caption{Stellar properties}
\label{tab:star}
\centering
\begin{tabular}{l c r}
\hline
\hline
\noalign{\smallskip}
\multicolumn{1}{l}{\multirow{4}*{Star names}} & \multicolumn{2}{c}{TOI-1055} \\
\multicolumn{1}{l}{}               & \multicolumn{2}{c}{HD\,183579} \\
\multicolumn{1}{l}{}               & \multicolumn{2}{c}{HIP\,96160} \\
\multicolumn{1}{l}{}               & \multicolumn{2}{c}{Gaia DR2 6641996571978861440} \\
\noalign{\smallskip}
\hline
\noalign{\smallskip}
Parameter & Value & Source \\
\noalign{\smallskip}
\hline
\noalign{\smallskip}
   $T_{\mathrm{eff}}$\; [K]         & $5784\pm61$       & Spectroscopy \\
   $\log{g}$                        & $4.45\pm0.10$     & Spectroscopy     \\\relax
   [Fe/H]                           & $-0.038\pm0.040$  & Spectroscopy    \\\relax
   [Mg/H]                           & $-0.06\pm0.01$  & Spectroscopy    \\\relax
   [Si/H]                           & $-0.06\pm0.03$  & Spectroscopy    \\
   $\pi$\; [mas]                    & $17.584\pm0.016$ & Gaia EDR3\tablefootmark{(a)}    \\
   $\theta$\; [mas]                & $0.1576\pm0.0013$ & IRFM\tablefootmark{(b)} \\
   $R_{\star}$\; [$R_{\odot}$]      & $0.960\pm0.006$   & IRFM\tablefootmark{(b)}  \\
   $M_{\star}$\; [$M_{\odot}$]     & $0.996\pm0.049$   & Isochrones    \\
   $t_{\star}$\; [Gyr]             & $3.1\pm1.9$       & Isochrones \\
   $L_{\star}$\; [$L_{\odot}$]     & $0.925\pm0.041$   & $R_{\star}$ \& $T_{\mathrm{eff}}$ \\
   $\rho_{\star}$\; [$\rho_{\odot}$] & $1.126\pm0.059$ & $R_{\star}$ \& $M_{\star}$ \\
   $\log{R'_{\mathrm{HK}}}$ & $-4.84\pm0.03$  & Spectroscopy \\
   $P_{\mathrm{rot}}$\; [d] & $24.8_{-4.4}^{+5.4}$  & $\log{R'_{\mathrm{HK}}}$ \\
\noalign{\smallskip}
\hline
\end{tabular}
\tablefoot{\tablefoottext{a}{Correction from \citet{lindegren2021} applied.}
\tablefoottext{b}{Infrared Flux Method.}}
\end{table}

\section{Methods}
\label{sec:method}
\subsection{Radial velocity extraction and detrending}
To unveil possible Keplerian signals induced by the orbital motion of planets, it is essential to properly model stellar activity in RV time series \citep[e.g.][]{saar1997,queloz2001,boisse2011,davis2017,dumusque2018,reiners2018,zhao2020,faria2020}. Stellar activity in Doppler measurements is often constrained by the change in the width and asymmetry of the CCFs used to measure the RVs of a star from high-resolution spectra \citep[e.g.][]{hatzes1996,queloz2001,queloz2009,figueira2013,simola2019}.

Following \citet{simola2019}, we modelled the CCFs as skew normal (SN) functions, which contain not only a location and a scale parameter (i.e. the counterparts of the mean and standard deviation of a normal function), but also a skewness parameter (hereafter denoted as $\gamma$) that quantifies the asymmetry of the function. In detail, we performed the SN fit to each available CCFs, retrieving the median $\overline{RV}$, the full width at half maximum, FWHM$_{\mathrm{SN}}$, the contrast, $A$, and the asymmetry, $\gamma$, of each SN function. To consistently attribute the uncertainties to each RV data point, $\sigma_{\mathrm{RV}}$, we followed a bootstrap approach after perturbing each CCF. For each epoch (i.e. for a given CCF), we perturbed the CCF data points by sampling values from a Normal distribution with mean $\mu=f_{\mathrm{CCF}}$ and standard deviation $\sigma=\sqrt{f_{\mathrm{CCF}}}$, where $f_{\mathrm{CCF}}$ is the CCF of that data point. After performing a SN fit onto each perturbed CCF, we retrieved the root mean square of the $\overline{RV}$ posterior distribution, which corresponds to $\sigma_{\mathrm{RV}}$.

The set $(\mathrm{FWHM_{SN}}, A, \gamma)$ constitutes the basis vector against which to detrend the RV measurements. To account for stellar activity, we applied the breakpoint (\emph{bp}) algorithm, which has been presented and broadly discussed in \citet{simola2022}. Here we briefly recall that the \emph{bp} method belongs to the class of Change Point Detection (CPD) techniques \citep[see, e.g.][and references therein]{page1954,truong2020,simola2022}, which detect the locations in a time series where correlation changes against the parameter of interest (in our case: FWHM$_{\mathrm{SN}}$, $A$, and $\gamma$) are statistically significant. The \emph{bp} method determines the optimal segmentation (i.e. the number of segments and their locations along a time series) for which the stellar activity does not change significantly within each segment, but rather between segments. \citet{simola2022} demonstrated that the RV detrending applied to each piecewise stationary segment of the RV time series removes the stellar activity contribution significantly better than the overall detrending applied on whole time series.

The \emph{bp} algorithm detected one change point location at the 12$^{\mathrm{th}}$ observation (BJD\,$\sim$\,2456165.6 corresponding to 26 August 2012, that is, within the HARPS-pre dataset), dividing the HARPS-pre RV measurements in two piecewise stationary segments. Since the change in the coefficients of correlation of RV against $(\mathrm{FWHM_{SN}}, A, \gamma)$ is judged to be statistically significant ($\Delta\mathrm{BIC}\approx-8$ in favour of that change point location), that date likely marks an inflection point in the stellar activity cycle with which $(\mathrm{FWHM_{SN}}, A, \gamma)$ are supposed to be correlated.

The RV detrending baseline we applied to each segment is a polynomial of the following form
\begin{equation}
RV_{\star} = \beta_0 + \displaystyle\sum_{k=1}^{k_t}{\beta_{k,t} t^k} + \displaystyle\sum_{k=1}^{k_F}{\beta_{k,F} \mathrm{FWHM_{SN}}^k} + \displaystyle\sum_{k=1}^{k_A}{\beta_{k,A} A^k} + \displaystyle\sum_{k=1}^{k_{\gamma}}{\beta_{k,\gamma} \gamma^k},
\label{eq:RVactivity}
\end{equation}
where the $\beta$ parameters are the polynomial coefficients and $(k_t, k_F, k_A, k_{\gamma})$ is the vector of polynomial degrees to be attributed to the regression versus time $t$, FWHM$_{\mathrm{SN}}$, $A$, and $\gamma$, respectively, whose best values are determined according to the Bayesian information criterion \citep[BIC;][]{schwarz1978} minimisation criterion. The parameters extracted from the CCFs and the detrended RV time series are listed in Tab.~\ref{tab:RVdata}. Instead, Fig.~\ref{fig:detrCoeff} shows the distributions inferred a posteriori of the activity-related coefficients of detrending (i.e. $\beta_0$, $\beta_{\cdot,F}$, $\beta_{\cdot,A}$, and $\beta_{\cdot,\gamma}$) for the two different piecewise stationary segments found within the HARPS-pre dataset. The difference between each pair emphasises that stellar activity impacts the RV observations of the two segments differently.
\subsection{Joint photometric and RV analysis}\label{sec:jointAnalysis}
We analysed the five transit light curves (three from TESS and two from CHEOPS) and the HARPS RV time series jointly following an MCMC approach and using the MCMCI code \citep{bonfanti2020}. As the star was already robustly characterised using two different stellar evolutionary codes (Sect.~\ref{sec:star}), we switched off the isochrone placement interaction at each chain step to save computational time.

Gaussian priors were imposed on the stellar parameters $T_{\mathrm{eff}}$, [Fe/H], $M_{\star}$, and $R_{\star}$, which have been used to derive (1) the best-fitting quadratic limb darkening (LD) coefficients $u_1$ and $u_2$ for both TESS and CHEOPS bandpasses following interpolation within a grid of \textsc{atlas9} models\footnote{\url{http://kurucz.harvard.edu/grids.html}} using the code described by \citet{espinoza2015}; (2) the mean stellar density, $\rho_{\star}$, that supports transit model convergence via Kepler's third law. We adopted uniform priors spanning the whole physical domain for all the planetary jump parameters, namely, the transit depth d$F\equiv\left(\frac{R_p}{R_{\star}}\right)^2$, the impact parameter, $b$, the transit timing, $T_0$, the orbital period, $P$, and the RV semi-amplitude $K_\mathrm{b}$. Though set to be uniform, priors on $\sqrt{e}\cos{\omega}$ and $\sqrt{e}\cos{\omega}$ were used to impose an eccentricity upper limit of $e<0.5$. The specific parameterisations adopted by the code are discussed in detail in \citet{bonfanti2020}.

We evaluated the best detrending models (in terms of low-order polynomials) to be applied to the flux and RV residuals, that is the LCs and RV time series after subtracting the transit and the Keplerian models, respectively. To this end, we launched several MCMCI runs comprising 10,000 steps each, changed each time the polynomial orders, and looked for the minimum BIC to select the best fitting model. In the case of the TESS LCs, the PDCSAP flux does not correlate with any of the variables provided in the SPOC data product. The two LCs extracted from Sector 13 only require a first-order polynomial to detrend flux against time, $t$, while the baseline function to be applied to the TESS LC from Sector 27 is a scalar (i.e. flat line). Similarly, we evaluated the photometric baseline for each of the CHEOPS LCs and the RV baseline according to Equation~(\ref{eq:RVactivity}). The detrending models employed for the joint LC and RV analysis are listed in Table~\ref{tab:detrending}.

We launched a preliminary longer run to evaluate the possible rescaling of the photometric errors, as detailed in \citet{bonfanti2020}, and obtain reliable error bars on the fitted parameters. Finally, we performed the global MCMC analysis employing four chains of 300,000 steps each, adding two RV jitter terms to account for any instrumental noise not included in our nominal uncertainties and/or possible sources of stellar variability not removed by our method. The chain convergence was checked through the Gelman-Rubin test \citep{gelman1992}.

\section{Results}\label{sec:results}
\subsection{Planet parameters}

\begin{figure}
\centering
\includegraphics[width=\columnwidth]{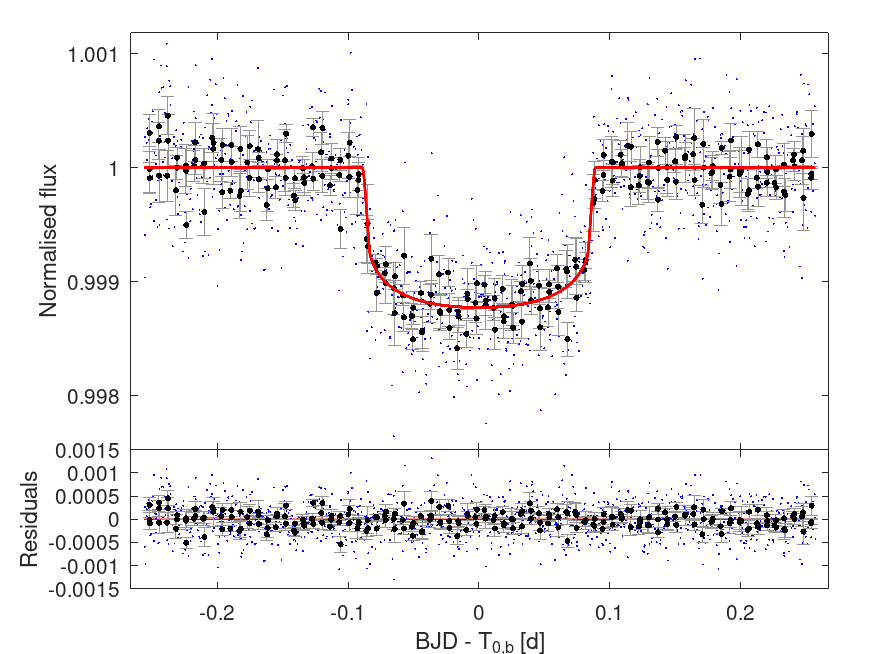} \\
\includegraphics[width=\columnwidth]{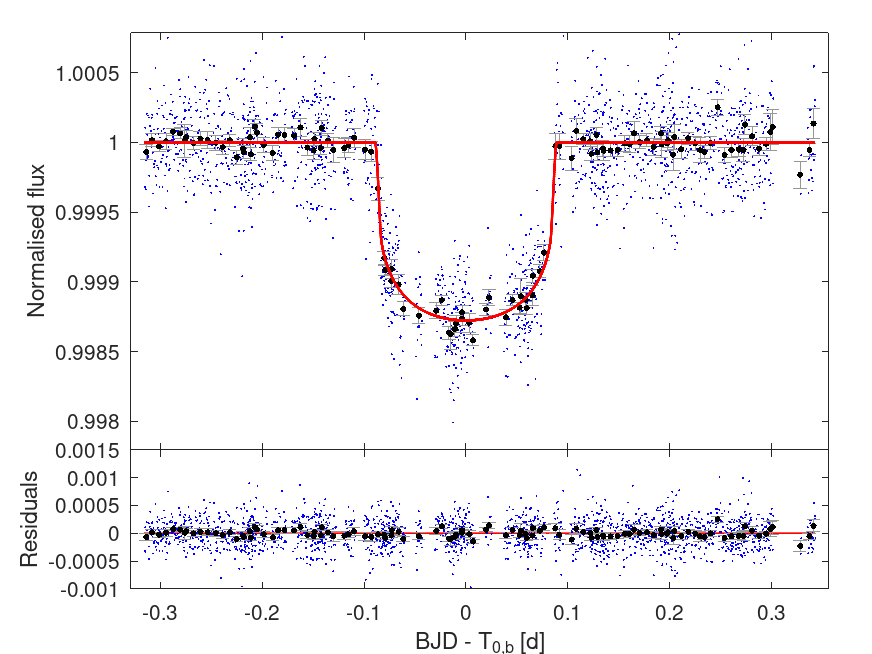}
\caption{Phase-folded LCs showing the transit of TOI-1055\,b from TESS (\textit{top panel}) and CHEOPS (\textit{bottom panel}). The red lines show the best-fit transit models.}
\label{fig:LCs}
\end{figure}

The MCMCI run successfully converged leading to the results reported in Table~\ref{tab:planet}. By refining the stellar parameters and adding two CHEOPS LCs to the already available three TESS LCs, we obtained a planetary radius of $R_\mathrm{b}=3.490_{-0.064}^{+0.070}$~$R_\oplus$. This is consistent with the radius measurement derived by P21 and G21, but has an improved precision of 1.9\%. Figure~\ref{fig:LCs} displays the phase-folded TESS and CHEOPS LCs.
As mentioned above, in our joint analysis we imposed Gaussian priors on both $M_{\star}$ and $R_{\star}$, which drive the convergence of the transit modelling through the constraint given by the consequent $\rho_{\star}$. To check whether our adopted stellar parameters are consistent with the photometric data, we further performed a run, where no a priori constraint was imposed on $\rho_{\star}$. From this transit modelling, we inferred $\rho_{\star,\mathrm{tr}}=1.84_{-0.49}^{+0.54}\,\rho_{\odot}$, which is consistent with our adopted $\rho_{\star}$ at the 1.4$\sigma$ level.

\begin{figure}
\centering
\includegraphics[width=\columnwidth]{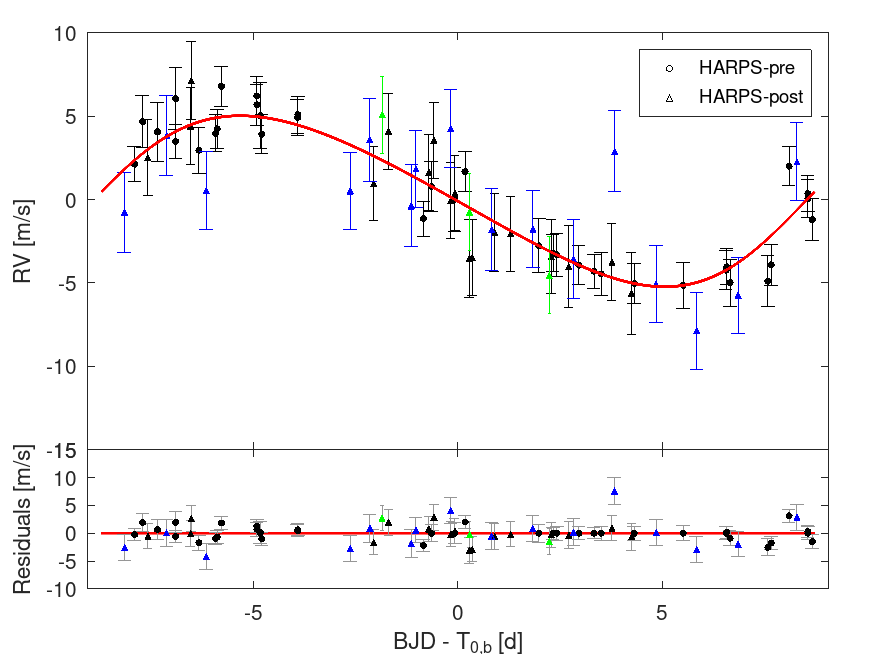}
\caption{Phase-folded RV signal produced by TOI-1055\,b with the best-fit Keplerian model superimposed (red). Marker colours distinguish the subsample used by P21 (black), the further three data points used by G21 (green), and the additional 16 data points we acquired. Instead, the marker shapes distinguish the measurement acquired before (HARPS-pre) and after (HARPS-post) the instrument upgrade. Error bars include the estimated jitter contributions.}
\label{fig:RV}
\end{figure}

From the RV analysis, we obtained $K_b=5.03_{-0.59}^{+0.60}$ m\,s$^{-1}$ (detection at $\sim$\,8$\sigma$ level), which implies a planetary mass of $M_\mathrm{b}=20.4_{-2.5}^{+2.6}\,M_{\oplus}$ and a mean density of $\rho_\mathrm{b}=2.65_{-0.35}^{+0.37}$ g\,cm$^{-3}$. Figure~\ref{fig:RV} shows the phase-folded RV curve of TOI-1055\,b. 
The SN-fit-based extraction of RV data, the application of the breakpoint analysis, and the additional 16 HARPS spectra enabled us to obtain a firm planetary mass measurement, with a significantly increased precision when comparing our results with the literature (see Table~\ref{tab:planetComparison}).

\begin{table}
\caption{Comparison of the planetary radius, $R_b$, mass, $M_b$, RV semi-amplitude, $K_b$, and average RV uncertainties accounting for the jitter, $\sigma_{\mathrm{RV+jitter}}$, as derived across the range of studies considered in this work.}
\label{tab:planetComparison}
\centering
\resizebox{\hsize}{!}{
\begin{tabular}{l c c c}
\hline\hline
\noalign{\smallskip}
Parameter & P21 & G21 & This work \\
\noalign{\smallskip}
\hline
\noalign{\smallskip}
$R_b$ [$R_{\oplus}$] & $3.55_{-0.12}^{+0.15}$ & $3.53_{-0.11}^{+0.13}$ & $3.490_{-0.064}^{+0.070}$ \\
\noalign{\smallskip}
$M_b$ [$M_{\oplus}$] & $19.7_{-3.9}^{+4.0}$ & $11.2\pm5.4$  & $20.4_{-2.5}^{+2.6}$ \\
\noalign{\smallskip}
$K_b$ [m\,s$^{-1}$] & $4.9_{-1.0}^{+0.9}$ & $2.7\pm1.3$     & $5.03_{-0.59}^{+0.60}$ \\
\noalign{\smallskip}
HARPS-pre $\sigma_{\mathrm{RV+jitter}}$ [m\,s$^{-1}$] & 2.86 & 5.29 & 1.29 \\
HARPS-post $\sigma_{\mathrm{RV+jitter}}$ [m\,s$^{-1}$] & 3.77 & 5.76 & 2.33 \\ 
\noalign{\smallskip}
\hline
\end{tabular}
}
\end{table}

We also performed the analysis with a more standard approach based on directly taking the data as released from the HARPS Data Reduction Software (DRS) and applying a polynomial detrending within the MCMC framework. The results agree well within 1$\sigma$ with those obtained from the SN-fit and \emph{bp}-based analysis, but the $K_b$ detection is at the $\sim$\,7$\sigma$ level (compared to the $\sim$\,8$\sigma$ level obtained using our novel technique), with $K_{b,\mathrm{DRS}}=5.05_{-0.69}^{+0.71}$\,m\,s$^{-1}$. Furthermore, after accounting for the jitter terms, the RV average error bars are larger than those obtained from our reference analysis by a factor of $\sim$\,2 and $\sim$1.2, respectively, for the HARPS-pre and HARPS-post datasets.

To further test our novel technique, instead of summing up all the order-by-order individual CCFs to get one global CCF per epoch, we built two different CCFs per epoch (CCF$_{\mathrm{blue}}$ and CCF$_{\mathrm{red}}$) by summing up all those individual CCFs derived from the first and the last 44 orders of the spectrograph, respectively. Then we extracted two different RV values per epoch by applying a SN-fit onto the two series of CCF$_{\mathrm{blue}}$ and CCF$_{\mathrm{red}}$, obtaining two different RV time series ($RV_{\mathrm{blue}}(t)$ and $RV_{\mathrm{red}}(t)$, respectively). After performing two different joint analyses, as described in Sect.~\ref{sec:jointAnalysis}, but inputting the $RV_{\mathrm{blue}}(t)$ (resp. $RV_{\mathrm{red}}(t)$) time series, we obtained $K_{b,\mathrm{blue}}=4.94\pm0.86$\,m\,s$^{-1}$ (resp. $K_{b,\mathrm{red}}=4.73\pm0.88$\,m\,s$^{-1}$), which agree well within 1$\sigma$ with the $K_b$ value of our reference analysis. Moreover, the rms of the RV residuals for the HARPS-pre and HARPS-post data resulted to be $\mathrm{rms_{blue}}=(3.5; 2.7)$\,m\,s$^{-1}$ and $\mathrm{rms_{red}}=(2.5; 2.4)$\,m\,s$^{-1}$, where the higher $\mathrm{rms_{blue}}$ values agree with expectation for which the line-profile distortion caused by stellar spots is higher at bluer wavelengths \citep[e.g][]{jeffers2022}.

Following our SN-fit extraction and \emph{bp} analysis, the mass we obtained is consistent within 1$\sigma$ with that found by P21. Remarkably, both P21 and particularly G21 needed to significantly increase the RV error bars through the jitter terms to model the HARPS data. Accounting for the jitters terms, the average RV uncertainties to be attributed to the HARPS-pre and HARPS-post measurements are (2.86; 3.77) and (5.29; 5.76) m\,s$^{-1}$ for P21 and G21, respectively. Following our analysis, the average RV errors after accounting for jitter terms are sensibly lower, being (1.29; 2.33) m\,s$^{-1}$ for the HARPS-pre and HARPS-post datasets. Following the $\log{R'_{\mathrm{HK}}}$--jitter relation of \citet{Hojjatpanah2020}, we obtain a jitter value of $\sim$\,4\,m\,s$^{-1}$, which, given the large scatter of a few m\,s$^{-1}$ intrinsic to the relation, is consistent with our results.

\begin{table}
\caption{TOI-1055\,b parameters derived from the joint LC and RV MCMC fit.}
\label{tab:planet}
\centering
\begin{tabular}{ll c}
\hline\hline
\noalign{\smallskip}
Parameter & Unit & Value \\
\noalign{\smallskip}
\hline
\noalign{\smallskip}
$P$ & [d]               & $17.471291_{-0.000028}^{+0.000027}$ \\
\noalign{\smallskip}
$T_0$ & [BJD$_\mathrm{TDB}$]           & $8661.06263_{-0.00096}^{+0.00097}$ \\
\noalign{\smallskip}
d$F$ & [ppm]            & $1106_{-35}^{+42}$ \\
$R_b/R_{\star}$ & & $0.03326_{-0.00053}^{+0.00063}$ \\
$b$ &                   & $0.48_{-0.12}^{+0.11}$ \\
\noalign{\smallskip}
$W$ & [h]               & $4.306_{-0.029}^{+0.031}$ \\
\noalign{\smallskip}
$u_{1,\mathrm{TESS}}$ & & $0.300\pm0.010$ \\
$u_{2,\mathrm{TESS}}$ & & $0.2752\pm0.0057$ \\
$u_{1,\mathrm{CHEOPS}}$ & & $0.417\pm0.012$ \\
$u_{2,\mathrm{CHEOPS}}$ & & $0.2555\pm0.0080$ \\
\noalign{\smallskip}
$a$ & [AU]              & $0.1322_{-0.0022}^{+0.0021}$ \\
\noalign{\smallskip}
$i_b$ & [$^{\circ}$]    & $89.11_{-0.13}^{+0.20}$ \\
\noalign{\smallskip}
$e$ &                   & $0.061_{-0.042}^{+0.061}$ \\
\noalign{\smallskip}
$\omega$ & [$^{\circ}$] & $270_{-61}^{+70}$ \\
\noalign{\smallskip}
$K_b$ & [m\,s$^{-1}$]     & $5.03_{-0.59}^{+0.60}$ \\
HARPS-pre RV jitter & [m\,s$^{-1}$] & $0.85_{-0.12}^{+0.09}$ \\
\noalign{\smallskip}
HARPS-post RV jitter & [m\,s$^{-1}$] & $2.13_{-0.04}^{+0.08}$  \\
\noalign{\smallskip}
$T_{\mathrm{eq}}$\tablefootmark{(a)} & [K] & $752\pm10$ \\
\noalign{\smallskip}
$R_b$ & [$R_{\oplus}$]  & $3.490_{-0.064}^{+0.070}$ \\
\noalign{\smallskip}
$M_b$ & [$M_{\oplus}$]  & $20.4_{-2.5}^{+2.6}$ \\
\noalign{\smallskip}
$\rho_b$ & [g\,cm$^{-3}$] & $2.65_{-0.35}^{+0.37}$ \\
\noalign{\smallskip}
\hline
\end{tabular}
\tablefoot{\tablefoottext{a}{Assuming zero albedo}.}
\end{table}

\subsection{Planet interior modelling}

The exquisite CHEOPS photometry and HARPS Doppler measurements, along with our statistical treatment of the RVs allowed us to determine the planetary mean density with a precision better than 15\%, which enables meaningful planetary interior modelling \citep{otegi2020}. We applied a Bayesian inference scheme to the derived stellar and planetary parameters (reported in Tables \ref{tab:star} and \ref{tab:planet}) to further investigate the internal structure of TOI-1055\,b. The used method is described in detail by \citet{leleu2021}. In the following, we briefly recall the most important aspects of our modelling approach and present the results it yields.

Our Bayesian analysis is based on a forward model that is used to calculate the radius that a planet with a certain combination of internal structure parameters would have. This information can then be used to determine the likelihood of this specific combination of internal structure parameters given the stellar and planetary observables, in particular, the transit depth and the RV semi-amplitude. The used forward model assumes the planet to be spherically symmetric and to consist of four fully distinct layers, namely an iron core, a silicate mantle, a water layer, and a H/He envelope, where the latter is modelled separately following \citet{lopez&fortney2014} and does not influence the other three layers. We further assume that the composition of the planet, namely, its Si/Mg/Fe ratios are identical to those measured in the host star's atmosphere \citep[see e.g.][]{dorn2015, thiabaud2015}; however, we note that although the stellar and planetary composition are indeed correlated, the relation might not be one-to-one \citep{adibekyan2021}. 

For the Bayesian analysis itself, we need to specify priors for all internal structure parameters. We chose a uniform prior for the mass fractions of the iron core, silicate mantle, and water layer with respect to the non-gaseous part of the planet, with the additional constraints that these mass fractions need to add up to unity and that the water mass fraction cannot be higher than 50\% \citep{thiabaud2014, marboeuf2014}. For the gas mass fraction of the planet, we chose a prior that is uniform in log. We stress that the results of our interior modelling do depend on the chosen priors to some extent.

The resulting posterior distributions for the interior of TOI-1055\,b are summarised in Figure \ref{fig:intStructure}. The first two columns show the posterior distributions of the mass fractions of the inner iron core and the water layer. Together with the mass fraction of the silicate mantle (not pictured), these add up to 1. The next three columns depict the molar fractions of silicon and magnesium within the silicate mantle and iron within the inner iron core. In the last column, we show the posterior of the gas mass of the planet in Earth masses on a logarithmic scale. Looking at these results, we expect TOI-1055\,b to host a substantial gas envelope with a mass of $0.41^{+0.34}_{-0.20}$ M$_\oplus$ and a thickness of $1.05^{+0.30}_{-0.29}$ R$_\oplus$. Instead, the water mass fraction is unconstrained.

\begin{figure}
\centering
\includegraphics[width=\columnwidth]{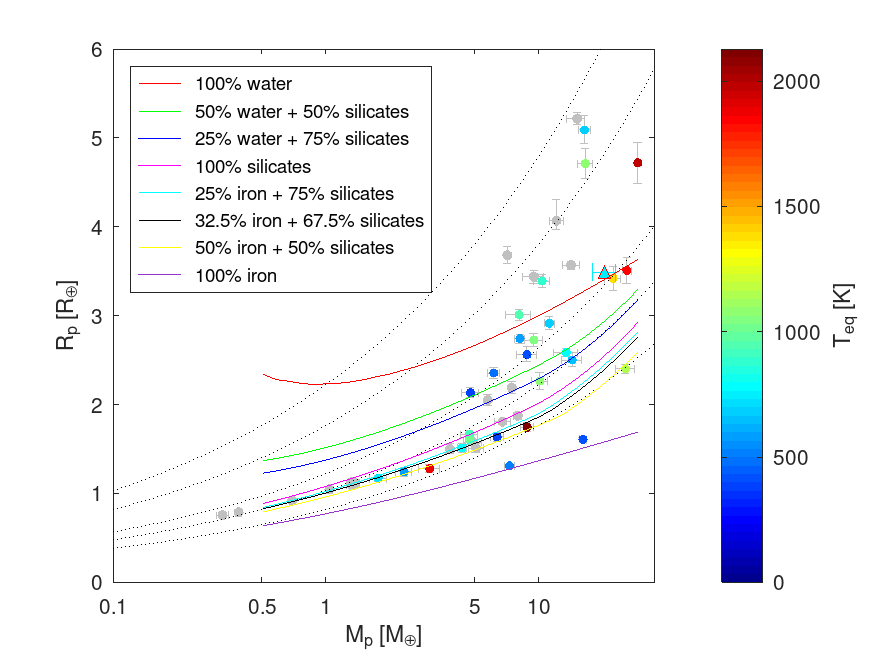}
\caption{Mass-radius diagram showing exoplanets with $M_p<30\,M_{\oplus}$, whose estimated mean density $\rho_{\rm p}$ is more precise than 15\%. TOI-1055\,b is represented by a triangle. All markers are colour-coded according to the planetary equilibrium temperature, $T_{\mathrm{eq}}$, where available. Dotted black lines represent the loci of constant $\rho_{\rm p}$ equal to 0.5, 1, 3, 5, and 10 g\,cm$^{-3}$ (from top to bottom). Solid lines represent different planet composition models (as explained in the legend) computed for an equilibrium temperature $T_{\mathrm{eq}}=T_{\mathrm{eq,TOI-1055b}}=750$ K.}
\label{fig:MRdiagram}
\end{figure}

Figure \ref{fig:MRdiagram} shows the position of TOI-1055\,b in a mass-radius diagram, together with different theoretical planet composition models. We remark that this diagram does not take into account the intrinsic degeneracy of modelling the internal structure of a planet. More specifically, even though TOI-1055\,b lies close to the theoretical model for a pure water sphere, other compositions, especially those including a H/He layer, can also reproduce the measured density. Furthermore, a planet made up of pure water is actually not considered to be physically realistic \citep[see e.g.][]{marboeuf2014}, which is also why we limit the water mass fraction in our internal structure model to an upper value of 50\%.

\section{Summary and conclusions}\label{sec:conclusions}

TOI-1055\,b is a transiting sub-Neptune for which two different mass estimates are available in the literature. In P21, 53 HARPS measurements were used to derive $K_{b,\mathrm{P21}}=4.9_{-1.0}^{+0.9}$ m\,s$^{-1}$, which implies $M_{b,\mathrm{P21}}=19.7_{-3.9}^{+4.0}\,M_{\oplus}$. G21 added three HARPS measurements to those considered by P21, obtaining $K_{b,\mathrm{G21}}=2.7\pm2.3$ m\,s$^{-1}$, that is, $M_{b,\mathrm{G21}}=11.2\pm5.4\,M_{\oplus}$. Given the statistical tension at the $\sim$\,2$\sigma$ level, we acquired further HARPS RV measurements and collected transit photometry with CHEOPS aiming at a more precise determination of the planetary mass and radius.

By performing a SN fit onto the HARPS CCFs, we extracted the RV time series together with $\mathrm{FWHM_{SN}}$, $A$, and $\gamma$, which enable RV detrending. We then applied the \emph{bp} algorithm \citep{simola2022} to detect the optimal locations along the time series, where the change in the correlations involving $\mathrm{FWHM_{SN}}$, $A$, and $\gamma$ are statistically significant. \citet{simola2022} showed that the stellar activity removal from the RV time series is much more effective when the detrending is separately applied to each piecewise stationary segment bounded by the breakpoint locations.

We jointly modelled five transit LCs (three from TESS and two from CHEOPS) and 72 HARPS measurements through the MCMCI code, simultaneously detrending the LCs and the RV time series. We derived $M_b=20.4_{-2.5}^{+2.6}\,M_{\oplus}$, $R_b=3.490_{-0.064}^{+0.070}\,R_{\oplus}$, and $\rho_b=2.65_{-0.35}^{+0.37}$ g\,cm$^{-3}$. The planetary mass is firmly detected (at $\sim$\,8$\sigma$) and agrees with the value derived by P21. According to our internal structure analysis, the planetary density can be explained by the presence of a substantial gas envelope made of H and He having a mass of $0.41^{+0.34}_{-0.20}$ M$_\oplus$ and a thickness of $1.05^{+0.30}_{-0.29}$ R$_\oplus$, while the water content is unconstrained.

The additional RV measurements and, particularly, the SN-fit-based RV extraction and the use of the \emph{bp} algorithm lead to a determination of the mass of TOI-1055\,b which is significantly more precise than the value previously published in the literature. Our results show that the SN fit combined with the \emph{bp} technique enables an efficient cleaning of the RV time series from stellar activity, providing a firm detection of the Doppler reflex motions induced by orbiting planets.

\begin{acknowledgements}
CHEOPS is an ESA mission in partnership with Switzerland with important contributions to the payload and the ground segment from Austria, Belgium, France, Germany, Hungary, Italy, Portugal, Spain, Sweden, and the United Kingdom. The CHEOPS Consortium would like to gratefully acknowledge the support received by all the agencies, offices, universities, and industries involved. Their flexibility and willingness to explore new approaches were essential to the success of this mission.
We thank the anonymous referee, whose valuable comments have improved the quality of the manuscript.
DG gratefully acknowledges financial support from the CRT foundation under Grant No. 2018.2323 ``Gaseousor rocky? Unveiling the nature of small worlds''. 
JAEg and YAl acknowledge the support of the Swiss National Fund under grant 200020\_172746. 
ML acknowledges support of the Swiss National Science Foundation under grant number PCEFP2\_194576. 
SS have received funding from the European Research Council (ERC) under the European Union’s Horizon 2020 research and innovation program (grant agreement No 833925, project STAREX). 
S.G.S. acknowledge support from FCT through FCT contract nr. CEECIND/00826/2018 and POPH/FSE (EC). 
TWi and ACCa acknowledge support from STFC consolidated grant numbers ST/R000824/1 and ST/V000861/1, and UKSA grant number ST/R003203/1. 
The Belgian participation to CHEOPS has been supported by the Belgian Federal Science Policy Office (BELSPO) in the framework of the PRODEX Program, and by the University of Liège through an ARC grant for Concerted Research Actions financed by the Wallonia-Brussels Federation. 
L.D. is an F.R.S.-FNRS Postdoctoral Researcher. 
This work has been carried out within the framework of the NCCR PlanetS supported by the Swiss National Science Foundation under grants 51NF40\_182901 and 51NF40\_205606. 
LMS gratefully acknowledges financial support from the CRT foundation under Grant No. 2018.2323 ‘Gaseous or rocky? Unveiling the nature of small worlds’. 
We acknowledge support from the Spanish Ministry of Science and Innovation and the European Regional Development Fund through grants ESP2016-80435-C2-1-R, ESP2016-80435-C2-2-R, PGC2018-098153-B-C33, PGC2018-098153-B-C31, ESP2017-87676-C5-1-R, MDM-2017-0737 Unidad de Excelencia Maria de Maeztu-Centro de Astrobiologí­a (INTA-CSIC), as well as the support of the Generalitat de Catalunya/CERCA programme. The MOC activities have been supported by the ESA contract No. 4000124370. 
S.C.C.B. acknowledges support from FCT through FCT contracts nr. IF/01312/2014/CP1215/CT0004. 
XB, SC, DG, MF and JL acknowledge their role as ESA-appointed CHEOPS science team members. 
LBo, VNa, IPa, GPi, RRa, and GSc acknowledge support from CHEOPS ASI-INAF agreement n. 2019-29-HH.0. 
ABr was supported by the SNSA. 
This project was supported by the CNES. 
This work was supported by FCT - Fundação para a Ciência e a Tecnologia through national funds and by FEDER through COMPETE2020 - Programa Operacional Competitividade e Internacionalizacão by these grants: UID/FIS/04434/2019, UIDB/04434/2020, UIDP/04434/2020, PTDC/FIS-AST/32113/2017 \& POCI-01-0145-FEDER- 032113, PTDC/FIS-AST/28953/2017 \& POCI-01-0145-FEDER-028953, PTDC/FIS-AST/28987/2017 \& POCI-01-0145-FEDER-028987, O.D.S.D. is supported in the form of work contract (DL 57/2016/CP1364/CT0004) funded by national funds through FCT. 
B.-O. D. acknowledges support from the Swiss State Secretariat for Education, Research and Innovation (SERI) under contract number MB22.00046. 
This project has received funding from the European Research Council (ERC) under the European Union’s Horizon 2020 research and innovation programme (project {\sc Four Aces}. 
grant agreement No 724427). It has also been carried out in the frame of the National Centre for Competence in Research PlanetS supported by the Swiss National Science Foundation (SNSF). DE acknowledges financial support from the Swiss National Science Foundation for project 200021\_200726. 
MF and CMP gratefully acknowledge the support of the Swedish National Space Agency (DNR 65/19, 174/18). 
M.G. is an F.R.S.-FNRS Senior Research Associate. 
SH gratefully acknowledges CNES funding through the grant 837319. 
KGI is the ESA CHEOPS Project Scientist and is responsible for the ESA CHEOPS Guest Observers Programme. She does not participate in, or contribute to, the definition of the Guaranteed Time Programme of the CHEOPS mission through which observations described in this paper have been taken, nor to any aspect of target selection for the programme. 
This work was granted access to the HPC resources of MesoPSL financed by the Region Ile de France and the project Equip@Meso (reference ANR-10-EQPX-29-01) of the programme Investissements d'Avenir supervised by the Agence Nationale pour la Recherche. 
PM acknowledges support from STFC research grant number ST/M001040/1. 
This work was also partially supported by a grant from the Simons Foundation (PI Queloz, grant number 327127). 
IRI acknowledges support from the Spanish Ministry of Science and Innovation and the European Regional Development Fund through grant PGC2018-098153-B- C33, as well as the support of the Generalitat de Catalunya/CERCA programme. 
GyMSz acknowledges the support of the Hungarian National Research, Development and Innovation Office (NKFIH) grant K-125015, a a PRODEX Experiment Agreement No. 4000137122, the Lend\"ulet LP2018-7/2021 grant of the Hungarian Academy of Science and the support of the city of Szombathely. 
V.V.G. is an F.R.S-FNRS Research Associate. 
NAW acknowledges UKSA grant ST/R004838/1.  

\end{acknowledgements}

\bibliographystyle{aa}
\bibliography{biblio}

\begin{appendix}

\section{Supplementary material}

\longtab[1]{ 
\begin{longtable}{rrrrrrrr}
\caption{\label{tab:RVdata} Radial velocities $\overline{RV}$ as extracted from the centred CCFs with their errors $\sigma_{\mathrm{RV}}$. They are followed by the hyperparameters inferred from the SN fit onto the CCFs (i.e. $\mathrm{FWHM_{SN}}$, $A$, and $\gamma$) and by the \emph{bp}-detrended RV values ($RV_{\mathrm{bp}}$) with their errors which also account for the jitter ($\sigma_{\mathrm{RV(bp+jitter)}}$).} \\
\hline\hline
\multicolumn{1}{c}{$\mathrm{BJD_{TDB}}$} & \multicolumn{1}{c}{$\overline{RV}$} & \multicolumn{1}{c}{$\sigma_{\mathrm{RV}}$} & \multicolumn{1}{c}{$\mathrm{FWHM_{SN}}$} & \multicolumn{1}{c}{$A$} & \multicolumn{1}{c}{$\gamma$} & \multicolumn{1}{c}{$RV_{\mathrm{bp}}$} & \multicolumn{1}{c}{$\sigma_{\mathrm{RV(bp+jitter)}}$} \\
\multicolumn{1}{c}{[$\mathrm{JD-2\,450\,000}$]} & \multicolumn{1}{c}{[m\,s$^{-1}$]} & \multicolumn{1}{c}{[m\,s$^{-1}$]} & \multicolumn{1}{c}{[km\,s$^{-1}$]} & \multicolumn{1}{c}{[\%]} & \multicolumn{1}{c}{} & \multicolumn{1}{c}{[m\,s$^{-1}$]} & \multicolumn{1}{c}{[m\,s$^{-1}$]} \\
\hline
\endfirsthead
\caption{continued.} \\
\hline\hline
\multicolumn{1}{c}{$\mathrm{BJD_{TDB}}$} & \multicolumn{1}{c}{$\overline{RV}$} & \multicolumn{1}{c}{$\sigma_{\mathrm{RV}}$} & \multicolumn{1}{c}{$\mathrm{FWHM_{SN}}$} & \multicolumn{1}{c}{$A$} & \multicolumn{1}{c}{$\gamma$} & \multicolumn{1}{c}{$RV_{\mathrm{bp}}$} & \multicolumn{1}{c}{$\sigma_{\mathrm{RV(bp+jitter)}}$} \\
\multicolumn{1}{c}{[$\mathrm{JD-2\,450\,000}$]} & \multicolumn{1}{c}{[m\,s$^{-1}$]} & \multicolumn{1}{c}{[m\,s$^{-1}$]} & \multicolumn{1}{c}{[km\,s$^{-1}$]} & \multicolumn{1}{c}{[\%]} & \multicolumn{1}{c}{} & \multicolumn{1}{c}{[m\,s$^{-1}$]} & \multicolumn{1}{c}{[m\,s$^{-1}$]} \\
\hline
\endhead
\hline
\endfoot
5847.53678513 &   9.743 & 1.250 & 7.092 & 33.320 & -0.0225 &  0.785 & 1.506 \\
5850.51602077 &   0.449 & 0.728 & 7.078 & 33.322 & -0.0134 & -3.207 & 1.112 \\
5851.51879988 &  -2.057 & 0.590 & 7.068 & 33.285 & -0.0157 & -4.300 & 1.027 \\
5852.50528724 &  -5.218 & 0.895 & 7.063 & 33.296 & -0.0182 & -5.031 & 1.227 \\
6042.79345732 &  -1.266 & 0.978 & 7.074 & 33.386 & -0.0224 & -3.318 & 1.289 \\
6043.87867695 &  -0.901 & 0.944 & 7.074 & 33.368 & -0.0188 & -4.461 & 1.264 \\
6045.88836933 &  10.194 & 1.104 & 7.079 & 33.356 & -0.0198 & -5.142 & 1.387 \\
6046.93991428 &   6.054 & 0.810 & 7.081 & 33.240 & -0.0160 & -4.228 & 1.167 \\
6048.94306496 &   4.349 & 0.786 & 7.075 & 33.052 & -0.0165 &  0.052 & 1.150 \\
6162.59232687 &   8.308 & 1.735 & 7.085 & 33.288 & -0.0138 &  0.003 & 1.928 \\
6164.65192835 &   3.057 & 1.376 & 7.081 & 33.285 & -0.0151 & -2.776 & 1.612 \\
6165.63332813 &   6.592 & 0.807 & 7.073 & 33.275 & -0.0182 & -3.930 & 1.165 \\
6378.90776269 & -11.627 & 0.669 & 7.073 & 33.270 & -0.0185 & -4.037 & 1.074 \\
6484.74450599 &  -8.916 & 1.302 & 7.090 & 33.312 & -0.0219 & -4.889 & 1.550 \\
6485.72460462 &  -6.043 & 0.714 & 7.082 & 33.334 & -0.0191 &  0.360 & 1.102 \\
6486.70288673 &  -3.610 & 0.654 & 7.084 & 33.315 & -0.0197 &  2.122 & 1.064 \\
6487.70666740 &  -2.297 & 0.606 & 7.082 & 33.316 & -0.0188 &  3.487 & 1.036 \\
6488.72760264 &  -1.224 & 0.789 & 7.088 & 33.329 & -0.0190 &  4.237 & 1.153 \\
6489.69485948 &   5.240 & 0.788 & 7.089 & 33.306 & -0.0144 &  6.208 & 1.152 \\
6490.70116321 &  -0.449 & 0.727 & 7.081 & 33.299 & -0.0168 &  5.090 & 1.111 \\
6557.59074155 &   2.601 & 1.650 & 7.088 & 33.228 & -0.0165 &  6.041 & 1.852 \\
6558.57259332 &  -1.161 & 0.725 & 7.074 & 33.267 & -0.0199 &  3.965 & 1.109 \\
6559.58347266 &   1.153 & 0.884 & 7.079 & 33.263 & -0.0208 &  5.686 & 1.219 \\
6560.57904039 &  -0.709 & 0.700 & 7.073 & 33.258 & -0.0204 &  4.913 & 1.093 \\
6850.71460802 &  -4.850 & 1.127 & 7.073 & 33.196 & -0.0159 & -4.977 & 1.405 \\
6851.72558860 &  -1.229 & 0.895 & 7.076 & 33.235 & -0.0146 & -3.913 & 1.227 \\
6852.73047310 &  -2.609 & 0.929 & 7.076 & 33.282 & -0.0189 & -1.210 & 1.253 \\
6853.79488843 &   2.774 & 1.319 & 7.088 & 33.288 & -0.0168 &  4.665 & 1.564 \\
6855.72584001 &   4.441 & 0.857 & 7.083 & 33.293 & -0.0184 &  6.793 & 1.200 \\
6856.71600423 &   1.386 & 0.795 & 7.085 & 33.288 & -0.0173 &  3.925 & 1.156 \\
6904.57466653 &   1.080 & 0.798 & 7.075 & 33.265 & -0.0200 &  2.004 & 1.159 \\
6906.57383255 &   5.307 & 1.536 & 7.079 & 33.213 & -0.0193 &  4.058 & 1.750 \\
6907.58764193 &   1.746 & 1.086 & 7.078 & 33.261 & -0.0221 &  2.956 & 1.373 \\
6961.50537101 &   8.572 & 1.770 & 7.086 & 33.265 & -0.0269 &  5.030 & 1.959 \\
6965.50200209 &  -7.177 & 0.625 & 7.083 & 33.300 & -0.0190 & -1.142 & 1.047 \\
6966.52198606 &  -4.859 & 0.821 & 7.079 & 33.309 & -0.0218 &  1.678 & 1.175 \\
7226.71719719 &   4.567 & 0.671 & 7.099 & 33.242 & -0.0085 &  4.092 & 2.281 \\
7227.70356856 &   3.339 & 0.617 & 7.102 & 33.246 & -0.0085 &  1.639 & 2.266 \\
7228.69831813 &  -1.179 & 0.847 & 7.102 & 33.252 & -0.0056 & -3.524 & 2.339 \\
7229.71155768 &  -0.510 & 0.533 & 7.098 & 33.244 & -0.0066 & -2.054 & 2.244 \\
7230.69925828 &  -3.856 & 0.540 & 7.094 & 33.223 & -0.0072 & -3.415 & 2.246 \\
7232.67077753 &  -2.385 & 1.085 & 7.103 & 33.241 & -0.0061 & -5.625 & 2.435 \\
7283.54965871 & -11.988 & 1.118 & 7.087 & 33.251 & -0.0088 & -4.030 & 2.450 \\
7284.60138328 & -13.061 & 0.766 & 7.090 & 33.278 & -0.0122 & -3.744 & 2.311 \\
7507.88041927 &  -8.615 & 0.640 & 7.116 & 33.303 & -0.0089 &  0.388 & 2.272 \\
7587.71855865 &  -4.987 & 0.724 & 7.086 & 33.250 & -0.0085 &  2.514 & 2.297 \\
7588.76805796 &  -4.577 & 0.780 & 7.085 & 33.265 & -0.0084 &  4.408 & 2.315 \\
7588.78034201 &  -4.781 & 0.780 & 7.081 & 33.231 & -0.0126 &  7.134 & 2.315 \\
7664.61178176 &  -1.244 & 0.652 & 7.093 & 33.318 & -0.0084 &  3.554 & 2.275 \\
7665.54986919 &  -6.840 & 0.642 & 7.095 & 33.304 & -0.0088 & -3.467 & 2.272 \\
7682.49965115 &  -7.995 & 0.670 & 7.090 & 33.275 & -0.0127 & -0.055 & 2.281 \\
7683.56929593 & -12.276 & 0.890 & 7.089 & 33.305 & -0.0150 & -1.965 & 2.355 \\
8047.50511052 &  -5.093 & 0.497 & 7.087 & 33.232 & -0.0105 &  0.967 & 2.236 \\
8711.61992092 &  10.641 & 0.745 & 7.099 & 33.220 & -0.0099 &  5.083 & 2.304 \\
8713.76108636 &   7.254 & 0.792 & 7.100 & 33.217 & -0.0081 & -0.744 & 2.319 \\
8715.71516874 &   1.617 & 0.788 & 7.102 & 33.241 & -0.0096 & -4.541 & 2.318 \\
9490.51985211 &   2.193 & 0.876 & 7.080 & 33.181 & -0.0077 &  2.268 & 2.350 \\
9491.51619503 &  -1.084 & 0.983 & 7.087 & 33.216 & -0.0101 & -0.778 & 2.391 \\
9492.54609652 &   3.925 & 1.001 & 7.077 & 33.164 & -0.0077 &  3.837 & 2.399 \\
9493.51875260 &   1.015 & 0.794 & 7.083 & 33.194 & -0.0080 &  0.543 & 2.320 \\
9497.52783204 &   7.482 & 1.173 & 7.092 & 33.239 & -0.0086 &  3.586 & 2.476 \\
9498.54318197 &   5.005 & 1.129 & 7.093 & 33.214 & -0.0094 & -0.379 & 2.455 \\
9499.50944480 &   7.827 & 0.906 & 7.089 & 33.200 & -0.0050 &  4.255 & 2.361 \\
9500.50980525 &   8.627 & 1.094 & 7.097 & 33.176 & -0.0051 & -1.805 & 2.439 \\
9501.52248169 &   8.608 & 0.830 & 7.095 & 33.169 & -0.0061 & -1.762 & 2.333 \\
9502.51735788 &   2.829 & 0.915 & 7.088 & 33.149 & -0.0062 & -3.585 & 2.364 \\
9503.52298688 &   7.672 & 1.061 & 7.089 & 33.189 & -0.0078 &  2.885 & 2.425 \\
9504.54147184 &   1.998 & 0.717 & 7.095 & 33.207 & -0.0048 & -5.062 & 2.295 \\
9505.53339626 &   0.510 & 0.744 & 7.091 & 33.154 & -0.0057 & -7.866 & 2.304 \\
9506.55052373 &  -3.067 & 0.652 & 7.085 & 33.154 & -0.0048 & -5.733 & 2.275 \\
9514.51711531 &  -1.677 & 0.712 & 7.083 & 33.206 & -0.0103 &  0.519 & 2.293 \\
9655.89641457 &   6.841 & 0.702 & 7.099 & 33.165 & -0.0031 &  1.839 & 2.290 \\
\end{longtable}

}

\begin{table}[h!]
\caption{Polynomial detrending baseline models. $c$ indicates a normalisation scalar; see text for the other symbols.}
\label{tab:detrending}
\centering
\begin{tabular}{l c}
\hline\hline
\noalign{\smallskip}
Time series & Detrending model \\
\noalign{\smallskip}
\hline
\noalign{\smallskip}
TESS1, Sector 13 & $t^1$ \\
TESS2, Sector 13 & $t^1$ \\
TESS1, Sector 27 & $c$ \\
CHEOPS \texttt{TG012301}\tablefootmark{(a)} & $t^1$ + \texttt{roll}$^2$ + \texttt{bg}$^3$ \\
CHEOPS \texttt{TG012302}\tablefootmark{(b)} & $t^4$ + \texttt{smear}$^1$ \\
HARPS-pre RV & $t^4$ + $\mathrm{FWHM_{SN}}^3$ + $\gamma^4$ + $A^1$ \\
HARPS-post RV & $t^3$ + $\mathrm{FWHM_{SN}}^3$ + $\gamma^4$ + $A^1$ \\
\noalign{\smallskip}
\hline
\end{tabular}
\tablefoot{\tablefoottext{a}{File key \texttt{CH\_PR100024\_TG012301\_V0200}}\\ \tablefoottext{b}{File key \texttt{CH\_PR100024\_TG012302\_V0200}}.}
\end{table}

\begin{figure}[h!]
\centering
\includegraphics[width=\columnwidth]{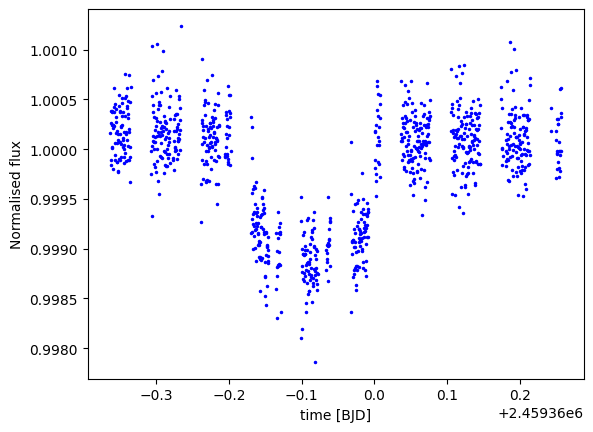}
\includegraphics[width=\columnwidth]{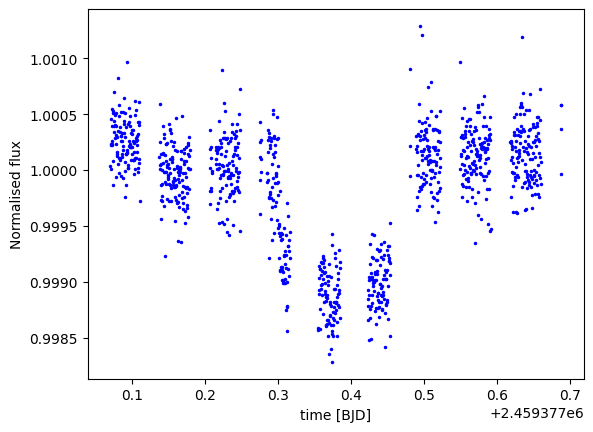}
\caption{Raw CHEOPS LCs displayed from top to bottom in chronological order of observation. \textit{Top panel:} File key \texttt{CH\_PR100024\_TG012301\_V0200}. \textit{Bottom panel:} File key \texttt{CH\_PR100024\_TG012302\_V0200}.}
\label{fig:rawCheopsLC}
\end{figure}

\begin{figure}[h!]
\centering
\includegraphics[width=\columnwidth]{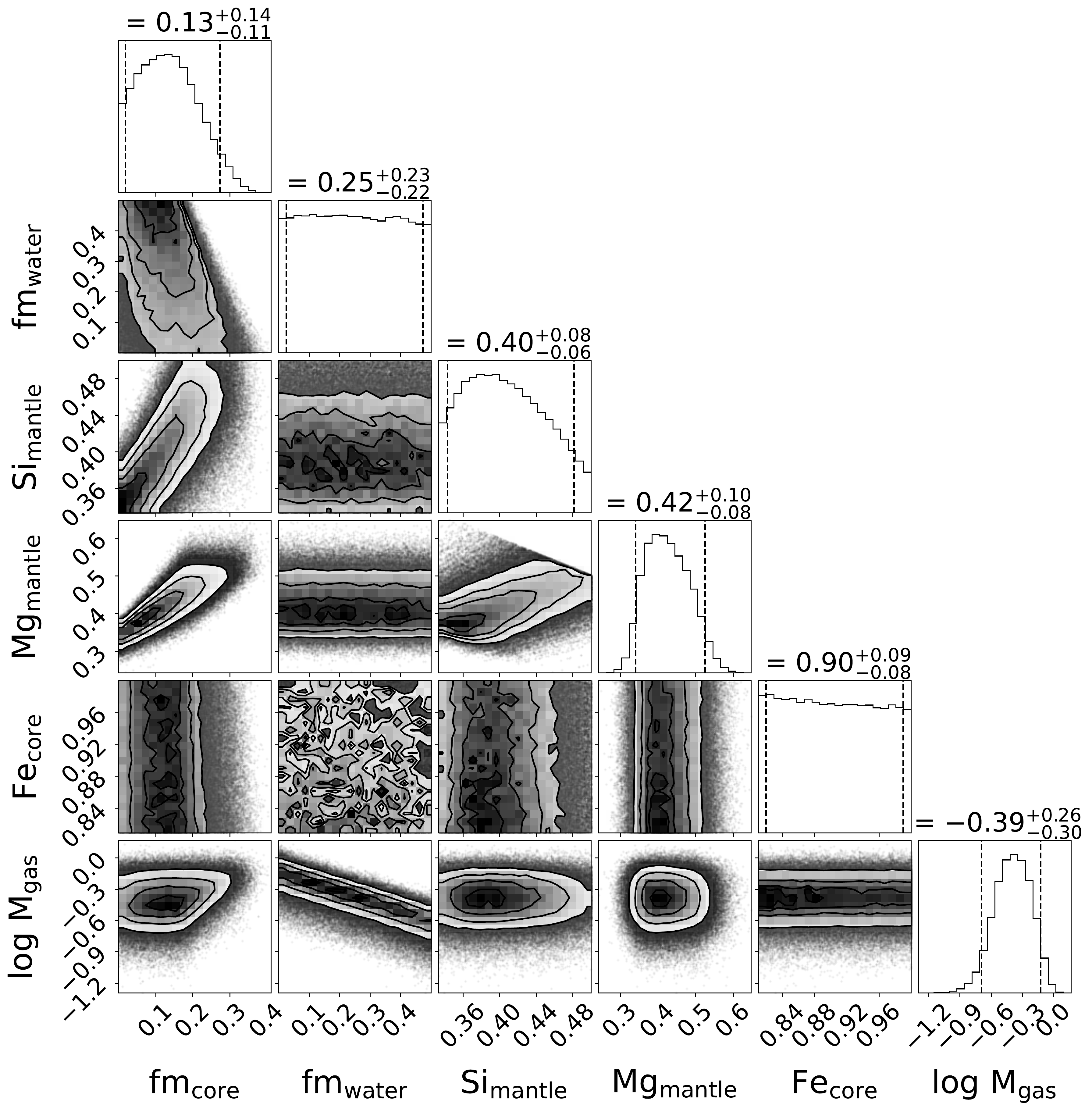}
\caption{Corner plot showing the posteriors of the most important internal structure parameters of TOI-1055\,b. Depicted are the distributions of the mass fractions of the inner iron core and the water layer with respect to the solid part of the planet, the molar fractions of Si and Mg in the mantle layer and of Fe in the inner core and the absolute mass of H/He gas in Earth masses, on a logarithmic scale. The titles of each column give the median and the 5 and 95 percentile of each posterior distribution.}
\label{fig:intStructure}
\end{figure}

\begin{figure*}
\centering
\includegraphics[width=0.33\textwidth]{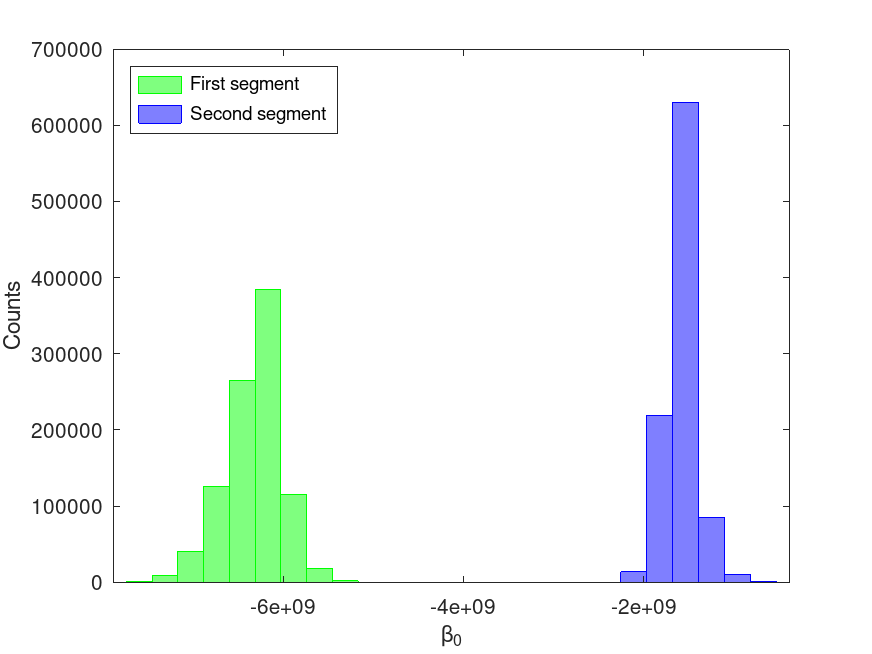}
\includegraphics[width=0.33\textwidth]{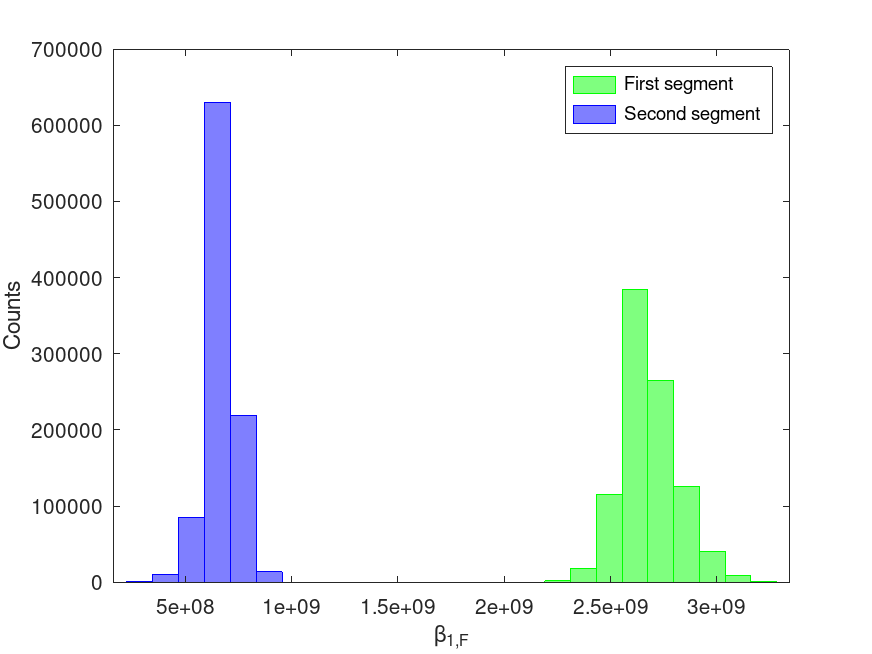}
\includegraphics[width=0.33\textwidth]{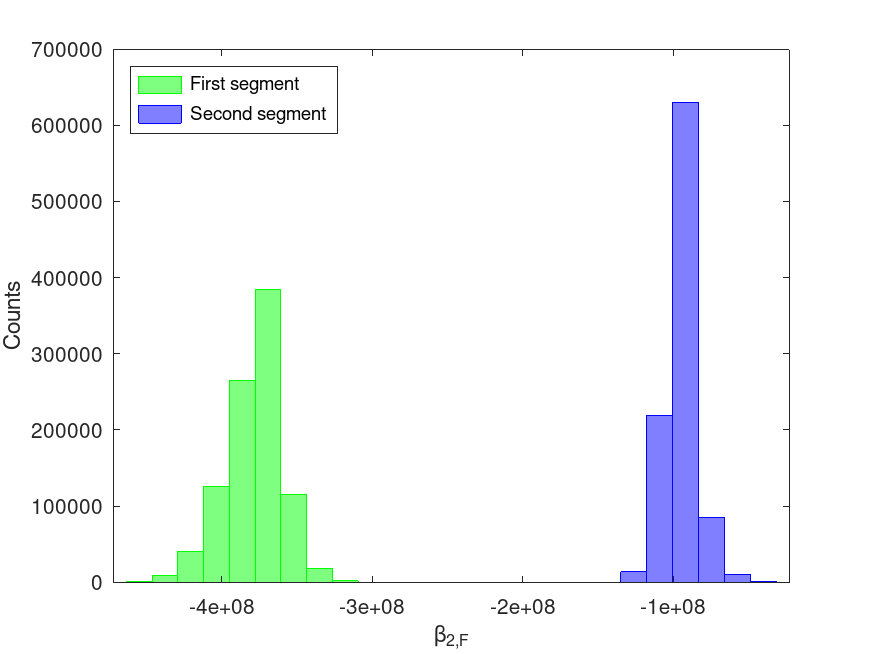}
\includegraphics[width=0.33\textwidth]{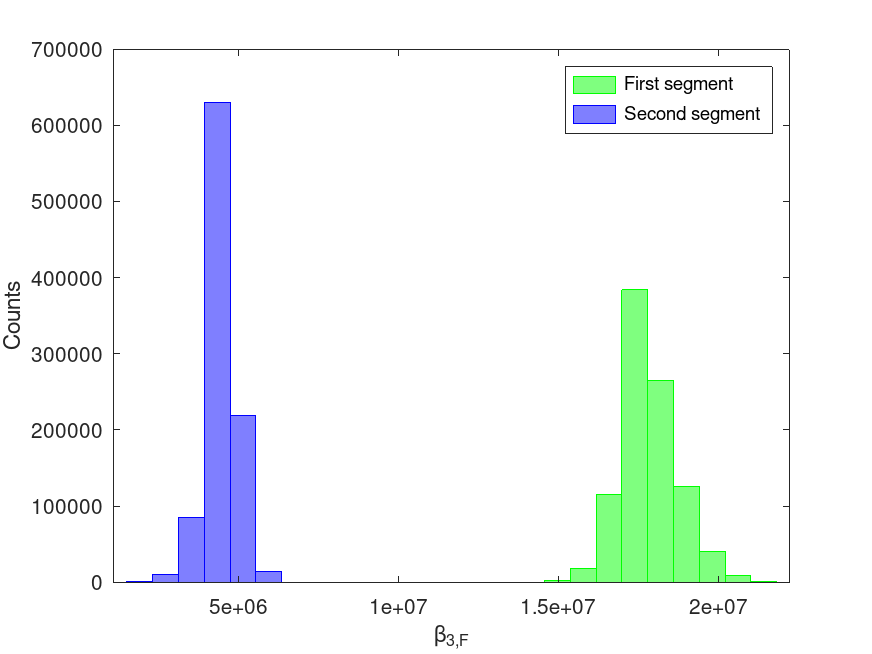}
\includegraphics[width=0.33\textwidth]{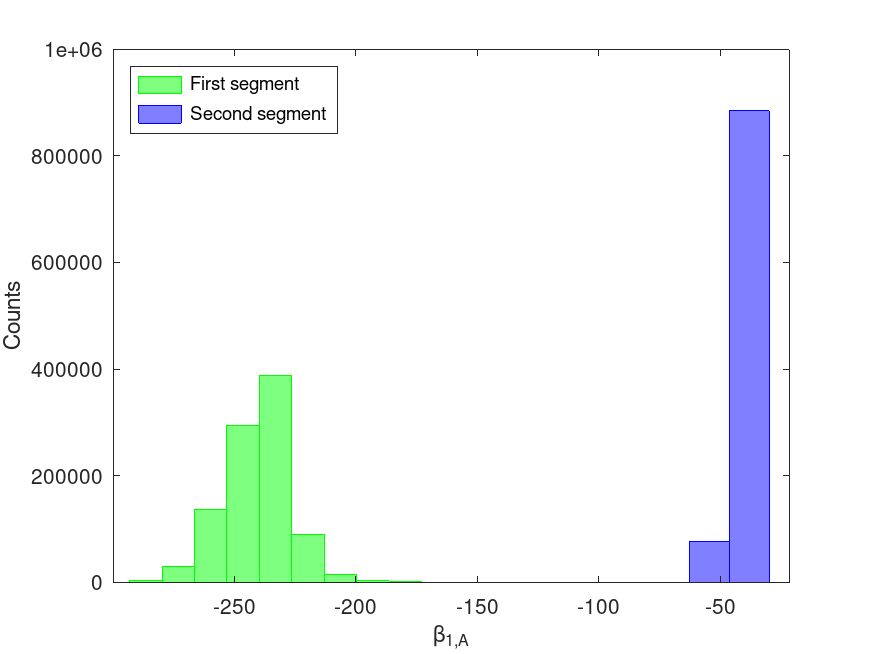}
\includegraphics[width=0.33\textwidth]{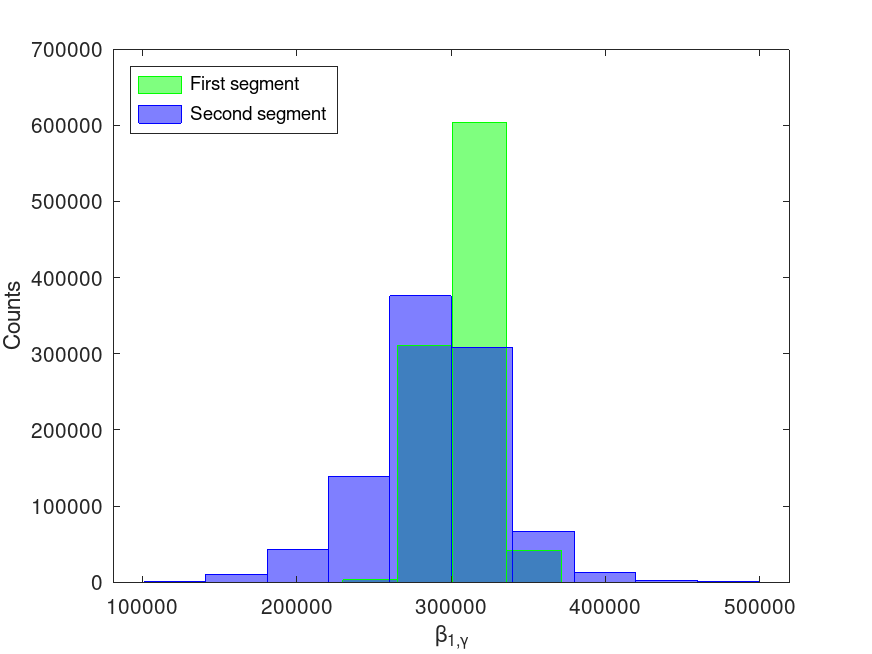}
\includegraphics[width=0.33\textwidth]{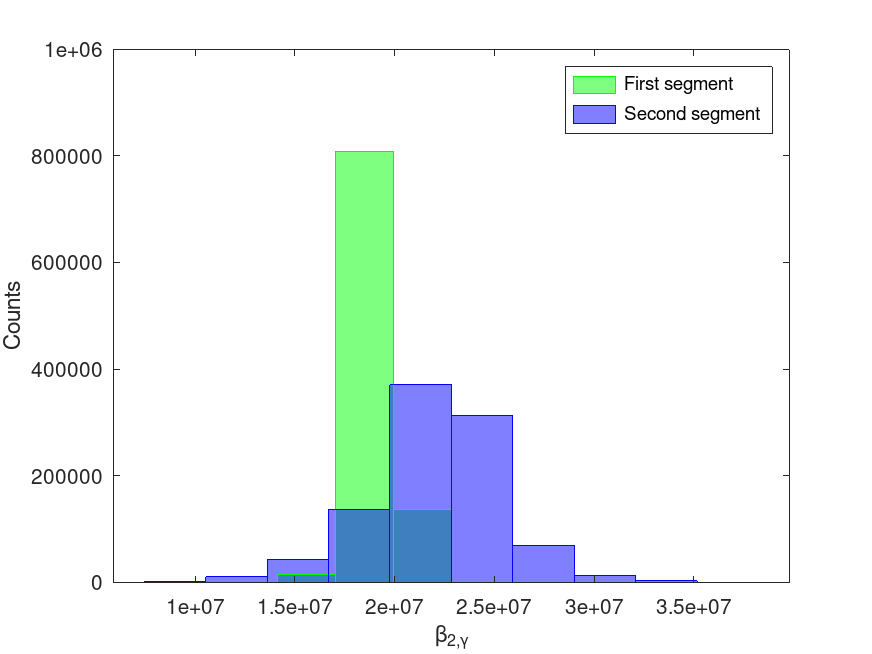}
\includegraphics[width=0.33\textwidth]{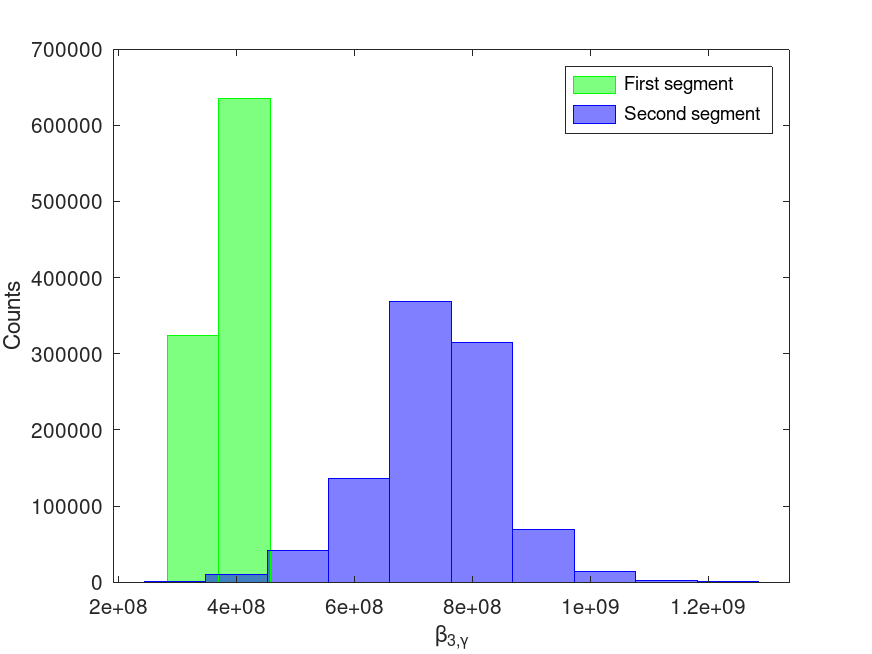}
\includegraphics[width=0.33\textwidth]{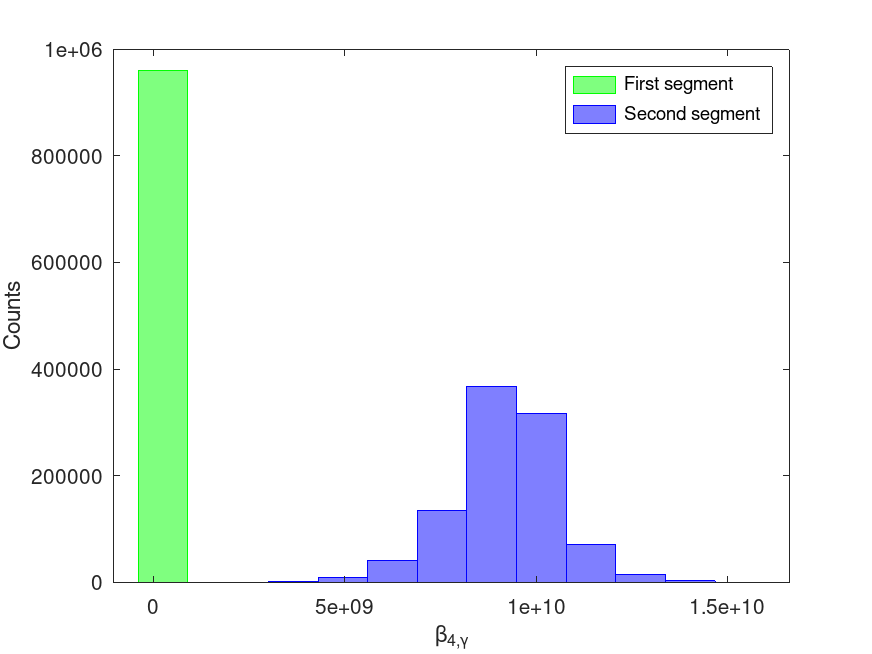}
\caption{Distributions inferred a posteriori for the activity-related detrending coefficients (i.e. $\beta_0$, $\beta_{\cdot,F}$, $\beta_{\cdot,A}$, and $\beta_{\cdot,\gamma}$ of Eq.~(\ref{eq:RVactivity}); the digit in the subscript refers to the polynomial order). Each panel compares the optimal detrending coefficients applied to the two different piecewise stationary segments that were found within the HARPS-pre dataset. The difference between the two distributions of each parameter emphasises that stellar activity impacts the RV observations differently when moving from one segment to the other.}
\label{fig:detrCoeff}
\end{figure*}

\end{appendix}

\end{document}